\documentclass[%
 reprint,
amsmath,amssymb,
aps,
prd,
]{revtex4-2}

\usepackage{graphicx}
\usepackage{dcolumn}
\usepackage{bm}
\usepackage{orcidlink}
\usepackage[caption=false]{subfig}
\usepackage{multirow}
\usepackage{bbold}
\usepackage[dvipsnames]{xcolor}
\usepackage{ulem}

\hypersetup{
	colorlinks=true,
	citecolor=NavyBlue,
	linkcolor=NavyBlue,
	urlcolor=NavyBlue,
}

\begin{document}

\title{Improved constraint on the Hubble constant from dark sirens with LIGO/Virgo/KAGRA O4a}

\author{Viviane Alfradique\orcidlink{0000-0002-5225-1923}}
\email{vivianeapa@cbpf.com}
\author{Clécio R. Bom\orcidlink{0000-0003-4383-2969}}
\author{Gabriel Teixeira\orcidlink{0000-0002-1594-208X}}
\author{André Santos\orcidlink{0000-0002-1420-3584}}
\affiliation{
 Centro Brasileiro de Pesquisas F\'isicas, Rua Dr. Xavier Sigaud 150, 
 22290-180 Rio de Janeiro, RJ, Brazil\\
}%

\date{\today}

\begin{abstract}
A new measurement of the Hubble constant $H_0$ is presented using the statistical dark siren method applied to a sample of seven well-localized gravitational-wave (GW) events from the fourth LIGO–Virgo–KAGRA (LVK) observing run and ten additional events from the first three runs. Galaxy catalogs from the DESI Legacy Imaging Survey (LS) are combined with a deep learning model to compute photometric redshift probability density functions. We extend our previous analysis by including the events GW230731\_215307 and GW230927\_153832, using sky maps from the fourth Gravitational-Wave Transient Catalog (GWTC-4), and introducing key methodological improvements: $r$-band luminosity weighting of host galaxies; an extended GW likelihood that incorporates information from the binary black hole component masses; and a consistent treatment of selection effects that accounts for the incompleteness of the magnitude-limited LS galaxy catalog. Using a total of 17 well-localized dark sirens (seven from the first part of the fourth observing run, O4a), we obtain $H_0 = 78.8^{+14.6}_{-12.2}\,\rm{km/s/Mpc}$ without luminosity weighting and $H_0 = 78.2^{+12.0}_{-11.0}\,\rm{km/s/Mpc}$ when applying $r$-band luminosity weighting. Finally, we combine the luminosity-weighted dark siren sample with the bright siren GW170817, including constraints on the jet viewing angle and corrections for the host galaxy peculiar velocity, to obtain a final constraint of $H_0 = 69.9^{+4.1}_{-4.0}\,\rm{km/s/Mpc}$, representing an improvement of approximately $11\%$ in the uncertainty relative to the GW170817-only result.
\end{abstract}

\maketitle

\section{Introduction}
\label{sec:intro}
Since its emergence, multimessenger astronomy has transformed both astrophysics and cosmology. The synergy between electromagnetic and gravitational wave observations, and potentially future neutrino detections \citep{Perego2014, Kimura2017, Fang2017, Biehl2018, Abbasi2023, Mukhopadhyay2024}, has opened the possibility to investigate phenomena from matter at extreme densities in neutron stars \citep{Janiuk2014, Kasen2017, Coughlin2019, Dietrich2020, Chen2024a} to precision measurements of cosmological parameters and tests of general relativity in a strong gravity field regime \citep{Mukhopadhyay2024, test_GR, Abbott2025}. Multimessenger observations have already placed constraints on the neutron star equation of state \citep{Radice2018, Abbott2018, Raithel2019}, improved our understanding of binary properties \citep{Margalit2017, Radice2019, Coughlin2019}, and contributed to a more comprehensive understanding of the physical processes involved in compact object mergers \citep{Kasen2017, Cowperthwaite2017, Kasliwal2017, Waxman2018}. 

Over the last decade, we have entered the era of precision cosmology, in which multiple observational probes constrain the parameters of the standard $\Lambda$CDM model with unprecedented accuracy. This growing precision has revealed considerable tensions between early- and late-time measurements of cosmological parameters, challenging and putting this model to the test. The most significant of these discrepancies is the so-called ``Hubble tension'' \citep{verde, DiValentino2021, abdalla2022cosmology}: a persistent $5$--$6\sigma$ disagreement between the value of the Hubble constant inferred from cosmic microwave background measurements by the Planck collaboration \citep{planck18} and the local distance-ladder determinations using Cepheid-calibrated supernovae from the SH0ES collaboration \citep{Breuval2024}. Additional tensions have also emerged, including $2$--$3\sigma$ discrepancies \citep{Hamana2020, Asgari2021, Dalal2023, Amon2024} involving the clustering parameter combination $S_{8} \equiv \sigma_{8}\sqrt{\Omega_{m0}/0.3}$, where $\sigma_{8}$ is the root mean square of the density contrast in spheres of radius $8\,h^{-1}$ Mpc and $\Omega_{m0}$ is the local matter density parameter. Furthermore, the DESI DR2 baryon acoustic oscillation measurements show a strong $2.8$--$4.2\sigma$ preference \citep{Lodha2025, Karim2025} for a dynamical dark energy component over the $\Lambda$CDM model.

In light of these tensions, GWs can serve as an independent cosmological probe, since they provide a direct measurement of the luminosity distance. Combined with the redshift of the host galaxy, they enable the construction of the Hubble diagram. This method, known as the ``standard siren'' approach \citep{schutz, 2005ApJ...629...15H, Dalal2006}, has become a powerful probe of cosmic expansion $H\left(z\right)$, especially since its first multi-messenger validation with GW170817 \citep{ligobns}. That event provided the first direct measurement of $H_0$ \citep{2017Natur.551...85A} and highlighted standard sirens as a promising tool for clarifying current cosmological tensions. When a merger produces a detectable electromagnetic counterpart, as in GW170817, in which an optical transient and gamma-ray burst enabled identification of the host galaxy NGC 4993 \citep{Coulter1556, marcelle17, valenti, arcavi, tanvir, lipunov}, the event is referred to as a bright siren.

An alternative way to use GWs to probe $H\left(z\right)$ is the statistical dark siren method, in which no electromagnetic counterpart is identified, and the galaxy distribution within the GW localization region is used to estimate the source redshift. Currently, two dark siren techniques have been validated with observational data. The first, the statistical dark siren method, assumes that each galaxy within the localization region is a potential host and assigns it a probability of hosting the event based on its position \citep{schutz, delpozzo, darksiren1, fishbach, palmese2023, Gray2023, Mastrogiovanni2023, alfradique2024, bom2024, gwtc4_cosmo, Pierra2026, McMahon2026}. The second approach relies on the spatial cross-correlation between the galaxy distribution and the population of GW mergers \citep{Namikawa2016, Mukherjee2021a, Mukherjee2021b, 2022MNRAS.511.2782C, Mukherjee2024, Ghosh2025, Matos2025}. Additionally, another class of method, using only GW data, aims to break the degeneracy between luminosity distance and redshift by exploiting features of the GW source parameters themselves, such as the intrinsic mass spectrum (the spectral siren method, \citep{Ezquiaga2021, Ezquiaga2022, Borghi2024, Magana2025, Farah2025, Tagliazucchi2025, Tagliazucchi2026, Pierra2026}), the spin distribution (spinning spectral sirens, \citep{Tong2025}), or, in the case of neutron star binaries, the binary-Love relations, which relate the tidal deformability to the source-frame mass through an EOS-dependent relation (the ``Love sirens'', \citep{Chatterjee2021, Ghosh2022}).

In this work, we focus on the statistical dark siren method to obtain a measurement of $H_0$, using the latest observations from the LVK O4a run combined with galaxy information from the public LS Survey catalog. We update the results presented in \citet{bom2024} by adding two new dark siren events to the sample and introducing several methodological improvements: (1) applying $r$-band luminosity weighting to each potential host galaxy; (2) using an extended GW likelihood that accounts for the dependence on BBH component masses; and (3) accounting for the incompleteness of the LS magnitude-limited galaxy catalog in the electromagnetic selection function. The analysis uses the sky maps released in GWTC-4 \citep{gwtc4_inference}, rather than the initial sky maps released shortly after the GW alerts. Throughout our analysis, we adopt a flat $\Lambda$CDM cosmology with a fixed matter density parameter, $\Omega_m = 0.3$, as our fiducial model.

\section{Data}\label{data}

\subsection{LIGO O4a run observation data}\label{sec:GWdata}
In this work, we revisit the five O4a GW events analyzed in \cite{bom2024} (GW230627\_015337, GW230919\_215712, GW230922\_020344, GW231206\_233901, and GW231226\_101520) and additionally include two new events, GW230927\_153832 and GW230731\_215307, resulting in a final sample of seven dark standard sirens from the O4a run. These events were selected to focus on well-localized sources, which are expected to provide the most informative posterior on $H_0$. The selection criteria adopted were the same as those used in previous works \citep{palmese2023, alfradique2024, bom2024}: (1) a luminosity distance $d_{L} < 1500$ Mpc, (2) a 50\% credible sky area $A_{50\%}<1000\, {\rm deg}^{2}$, and (3) at least 70\% probability coverage by the LS. The adopted cuts were propagated into the modeling of GW selection effects so that the dark siren $H_0$ inference remained fully consistent with the applied event selection.

Our selection criteria initially included GW230518$\_$125908 and GW230529$\_$181500. However, these events were excluded from the analysis. The first was detected during the engineering run (pre-O4) and was removed to avoid potential inconsistencies arising from non-standard detector conditions, which are not accounted for in our modeling of the GW selection effects. The second is consistent with a neutron star–black hole merger and was excluded to ensure consistency with this modeling, developed specifically for BBH events. For all GW events included in our analysis, we used posterior samples generated with the IMRPhenomXPHM\_SpinTaylor waveform model, as presented in \cite{posteriorO4a}. 

The information for our dark siren sample is reported in Table \ref{tab:events}. The sky coverage of the photometric redshift (photo-$z$) catalog for each GW event was estimated using the Multi-Order Coverage (MOC) maps defined by the International Virtual Observatory Alliance and implemented through the {\tt mocpy} Python library \citep{moc}. MOCs provide a compact representation of arbitrary sky regions based on the hierarchical equal-area pixelization scheme of Hierarchical Equal Area isoLatitude Pixelation of a sphere (HEALPix, \citep{healpix}). In this framework, sky regions are encoded as sets of HEALPix cells across multiple orders, enabling efficient spatial operations via integer-based set algebra rather than special polygon clipping. To estimate the coverage, we first filtered the LS catalog by converting photo-$z$ into luminosity distances under a fiducial cosmology consistent with the Planck 2015 constraints \citep{planck2015}. Galaxies within the luminosity distance range compatible with each GW event were retained as potential hosts. We then constructed an MOC representation of the spatial distribution of the selected galaxies using their sky coordinates. The MOC was generated at a maximum HEALPix order of 11, corresponding to an angular resolution of approximately 1.7 arcimnutes. At this order, each retained galaxy contributes its enclosing HEALPix cell, effectively approximating the angular footprint of the photo-$z$ catalog at the chosen resolution. Independently, the GW sky map was converted into a MOC by selecting HEALPix pixels comprising the 90\% CI, determined by ranking pixels in descending probability density and integrating until the cumulative probability reached 0.9. This region was likewise represented at maximum order 11 to ensure consistent angular resolution between datasets. The fractional photo-$z$ coverage of the GW localization area was computed by intersecting the galaxy MOC with the 90\% CI GW MOC. The sky area of the intersection $\left(\Omega_{\mathrm{photo\text{-}z} \cap \mathrm{GW},90\%}\right)$ and the area corresponding to the full GW credible region $\left(\Omega_{\rm{GW},90\%}\right)$ were calculated directly from the solid angles of ther constituent HEALPix cells using {\tt mocpy}. The coverage fraction is therefore defined as
\begin{align}
    f_{\mbox{cov}} = \frac{\Omega_{\text{photo-$z$} \cap  \rm{GW},90\%}}{\Omega_{\rm{GW},90\%}},
\end{align}
where the sky areas (in steradians) are derived from the MOC representations. This procedure ensures that the coverage estimate is resolution-controlled, computationally efficient, and fully consistent with the probabilistic structure of the GW localization. Three events have less than 90\% coverage of their credible regions, with two of them lying exactly at the threshold of our adopted selection criterion. For these events, we injected fake galaxies (following the same procedure used in our previous work, \citealt{bom2024}; see the description in Section  \ref{sec:fake}) into the uncovered regions to reproduce the completeness correction commonly applied in dark siren analyses \citep{chen17, fishbach, Finke2021, Gray2023a, posteriorO4a}. In the last column of Table \ref{tab:events}, we present the number of galaxies within the GW sky localization. The number inside the parentheses corresponds to the number of galaxies remaining after applying the apparent magnitude (survey sensitivity) cut, the absolute magnitude cut (to ensure a volume-limited sample, as detailed in the next subsection), and the additional quality cuts described in the following subsections. The number outside the parentheses indicates the total number of galaxies before any cuts are applied.

\begin{table*}
\caption{Summary of the properties of the O4a dark siren events analyzed in this work. Columns list, in order: the GW event name; the primary and secondary BH masses; the luminosity distance; the 90\% credible interval (CI) sky area and volume; the false alarm rate; the sky map coverage (in percent); and the number of potential host galaxies ($N_{\rm gal, host}$). The measurements are derived from posterior samples generated with the IMRPhenomXPHM\_SpinTaylor waveform model, as provided by \citet{posteriorO4a}, while the sky map coverage and $N_{\rm gal,host}$ are computed as described in the text.}
\centering
\begin{tabular}{ccccccccc}
\hline \hline
\multirow{2}{*}{O4a GW event} & \multirow{2}{*}{$m_1^{\rm det}\, [M_\odot]$} & \multirow{2}{*}{$m_2^{\rm det}\, [M_\odot]$} & \multirow{2}{*}{$d_L$ [Mpc]} & \multirow{2}{*}{A [deg$^2$]} & \multirow{2}{*}{V [Gpc$^3$]} & \multirow{2}{*}{FAR [yr$^{-1}$]} & Sky map & \multirow{2}{*}{$N_{\rm gal, host}$}\\
& & & & & & & Coverage [\%] & \\
\hline \hline
GW230627\_015337 & $9.0^{+1.9}_{-1.3}$ & $6.0^{+1.0}_{-0.9}$ & $307^{+62}_{-131}$ & 92 & $3.3 \times 10^{-4}$ &  $<1.0 \times 10^{-5}$ & 90  & 3103868(16076)\\
GW230731\_215307 & $12.0^{+3.0}_{-1.4}$ & $10.0^{+1.2}_{-1.9}$ & $1113^{+335}_{-453}$ & 629 & $8 \times 10^{-2}$ &  $<1.0 \times 10^{-5}$ & 73 & 17640655(26348)\\
GW230919\_215712 & $34^{+7}_{-4}$ & $27^{+3}_{-5}$ & $1275^{+734}_{-483}$ & 570 & $1.7 \times 10^{-1}$ & $<1.0 \times 10^{-5}$ & 70 & 17290478(27576) \\
GW230922\_020344 & $53^{+12}_{-9}$ & $37^{+7}_{-8}$ & $1453^{+763}_{-609}$ & 270 & $1 \times 10^{-1}$ & $3.6 \times 10^{-4}$ & 98 & 9032150(5251)\\
GW230927\_153832 & $27^{+4}_{-3}$ & $20^{+2.8}_{-2.7}$ & $1168^{+385}_{-516}$ & 273 & $4 \times 10^{-2}$ & $<1.0 \times 10^{-5}$ & 70 & 6235681(5880)\\
GW231206\_233901 & $48^{+9}_{-5}$ & $37^{+6}_{-8}$ & $1483^{+333}_{-492}$ & 292 & $6 \times 10^{-2}$ & $1.6 \times 10^{-5}$ & 99 & 10899526(19288)\\
GW231226\_101520 & $49^{+7}_{-3}$ & $43^{+3}_{-6}$ & $1168^{+222}_{-301}$ & 136 & $1 \times 10^{-2}$ & $<1.0 \times 10^{-5}$ & 98 & 4755915(11367) \\  
\hline \hline
\end{tabular}
\label{tab:events}
\end{table*}

\subsection{DESI Legacy Survey data}\label{sec:desi}

\begin{figure}
    \centering
    \includegraphics[width=\linewidth]{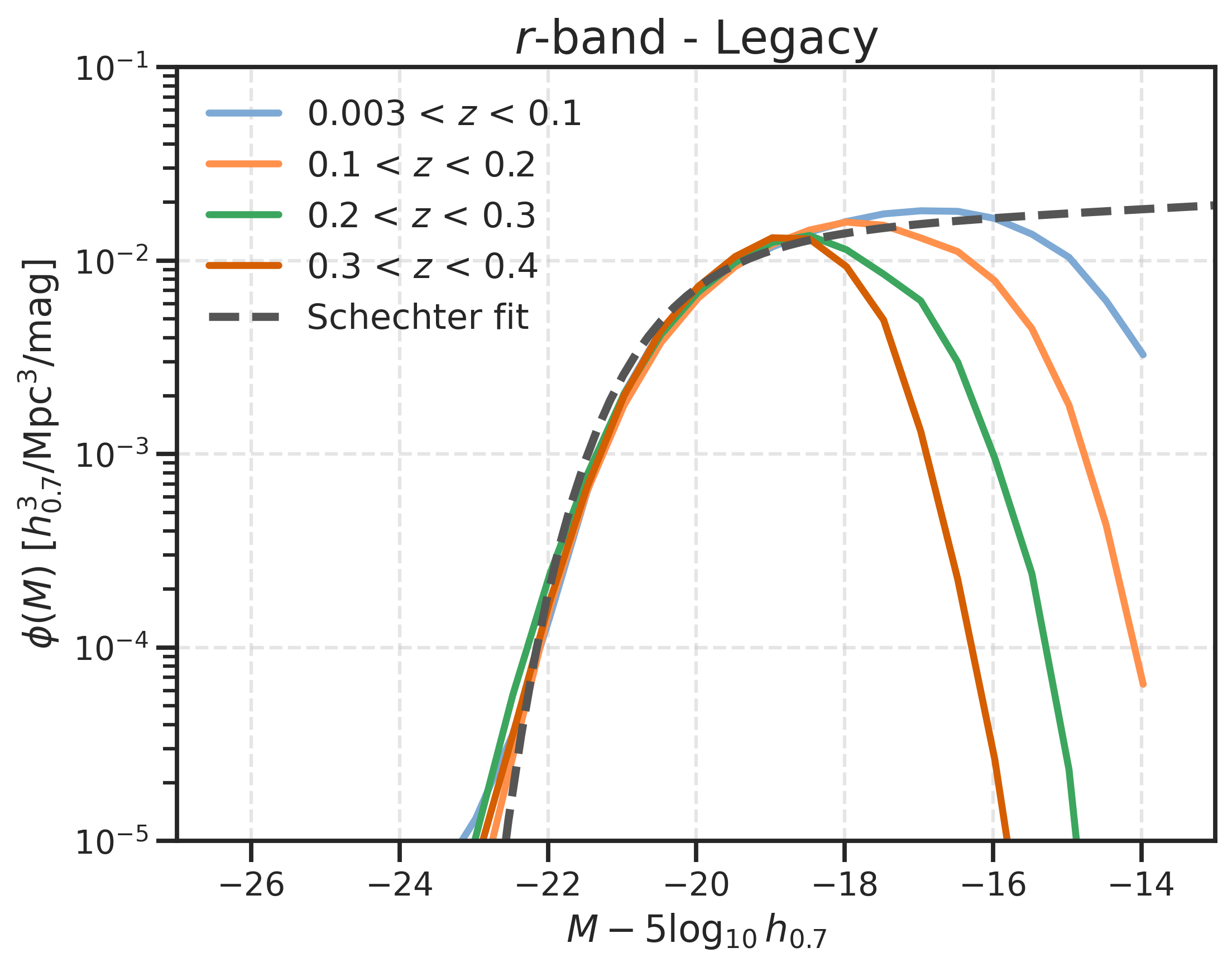}
    \caption{Schechter-function fit (gray) to the galaxy luminosity distributions of the LS in the \textit{r}-band. Different colors correspond to different redshift bins.}
\label{fig:schechter_function_Rband}
\end{figure}

The DESI Legacy Imaging Survey \citep{dey2019} is a combination of three public imaging surveys: the Dark Energy Camera Legacy Survey \citep{Blum2016}, the Beijing-Arizona Sky Survey \citep{Zou2017}, and the Mayall \textit{z}-band Legacy Survey \citep{Silva2016AAS}. Together, these surveys image a large fraction of the extragalactic sky. Imaging of the southern and equatorial regions was carried out with the Dark Energy Camera (DECam \citep{Flaugher2015}) as part of the DECam Legacy Survey (DECaLS), while the northern footprint, at declinations $\delta \gtrsim +32^\circ$, was imaged in \textit{gr} obtained with the Bok 2.3-m telescope as part of the Beijing-Arizona Sky Survey (BASS), and \textit{z}-band imaging acquired with the Mosaic-3 camera on the Mayall 4-m telescope as part of the Mayall \textit{z}-band Legacy Survey (MzLS). The photometric coverage was extended to include the \textit{i}-band via additional DECam observations, with most of the data originating from the full six years of Dark Energy Survey \citep{desdr2}, the DECam Local Volume Exploration Survey (DELVE \citep{delvedr2}), and the DECam eROSITA Survey (DeROSITAS \footnote{\href{https://noirlab.edu/science/programs/ctio/instruments/Dark-Energy-Camera/DeROSITAS}{https://noirlab.edu/science/programs/ctio/instruments/Dark-Energy-Camera/DeROSITAS}}). The imaging reaches typical 5$\sigma$ point-source depths of \textit{g} $\sim$ 24.0, \textit{r} $\sim$ 23.4, \textit{z} $\sim$ 22.5 AB mag. In this work, we primarily use the publicly available catalog from Data Release 10 (LS-DR10 \footnote{Data are available at \href{https://www.legacysurvey.org/dr10/}{https://www.legacysurvey.org/dr10.}}), which spans approximately 20,000 deg$^2$. For regions in the northern footprint, we instead use Data Release 9 (LS-DR9 \footnote{Data are available at \href{https://www.legacysurvey.org/dr9/}{https://www.legacysurvey.org/dr9.}}), which includes imaging from the BASS and the MzLS that are not included in DR10. To estimate the catalog completeness, we show that the LS \textit{r}-band absolute magnitude distribution is well described by a Schechter function adopting the SDSS-derived parameters \citep{Blanton2003}, $M_{*,r} = -20.44$, $\alpha = -1.05$, and $\phi_{*} = 1.49 \times 10^{-2}\,h^{3}\,\mathrm{Mpc}^{-3}$. Figure~\ref{fig:schechter_function_Rband} illustrates this behavior, showing good agreement at the bright end and the expected decline toward the survey threshold. Given this luminosity function and the adopted \textit{r}-band survey depth, we find that the LS achieves $\sim 99\%$ completeness at $z = 0.5$ for $M_r < -18.09$, and $\sim 98\%$ at $z = 0.6$ for $M_r < -18.56$.

To ensure that the resulting catalog is volume-limited, we follow the procedure described in previous work \citep{palmese20_sts,palmese2023,alfradique2024,bom2024}. First, for each dark siren event, we compute the maximum redshift corresponding to the upper limit of the 90\% CI on the luminosity distance, adopting the highest value of our $H_0$ prior. Given the apparent magnitude threshold at this maximum redshift, we then compute the faintest absolute magnitude that a galaxy could have and still satisfy the survey magnitude limit. All galaxies in the catalog with absolute magnitudes fainter than this threshold are removed. Throughout this procedure, we assume our fiducial $\Lambda$CDM cosmology.

\subsection{Photometric redshift estimation with Mixture Density Networks}
\label{sec:photoz}

The photometric redshifts used in this work were estimated using a deep learning model following the methodology presented in \citet{bom2024}. This method is based on mixture density networks (MDNs, \citep{Bishop_1994}), a technique that estimates probability density functions (PDFs) for the redshifts, represented as a mixture of Gaussians.

For galaxies with declination below 30 deg, we adopt the publicly available photometric data from the LS-DR10. For this subsample, we directly used the same MDN model presented in \citet{bom2024}, ensuring that our photo-$z$ estimates achieve the same level of performance reported in that work.

In contrast, for galaxies with declination above 30 deg, we adopt a different strategy using LS-DR9 data. Instead of using a model trained on the combined northern and southern samples, as in \citet{bom2024}, we train the MDN using objects exclusively from the northern region of the LS DR9 catalog. The training sample was constructed by first selecting sources with $\texttt{TYPE} \neq \texttt{PSF}$, to exclude objects classified as point sources by the LS detection pipeline \citep{dey2019}. We then restrict the sample to objects with available spectroscopic redshift measurements in the same catalog.

To ensure data quality, we excluded objects close to the detection limit by imposing $r < 22.5$, and removed sources with unphysical colors by requiring $-1 < g-r < 4$. These cuts are adapted from those used in \citet{bom2024}, although only a subset of their selection criteria was applied here--mostly due to the lack of $i$-band measurements for the northern galaxies. The remaining dataset was randomly split, reserving $20\%$ for testing in order to evaluate the performance of the network after training. Unlike the approach adopted in \citet{bom2024, alfradique2024, teixeira2024}, we did not homogenize the training distribution due to the limited number of reliable sources (approximately $40,000$ objects).

One of the main differences compared to \citet{bom2024} lies in the network architecture. Given the limited size of the training sample, we find that a simpler architecture yields better performance. The model consists of three fully connected layers with 16, 12, and 8 units, respectively, followed by an MDN layer with six Gaussian components.

\begin{figure}
    \centering
    \includegraphics[width=1\linewidth]{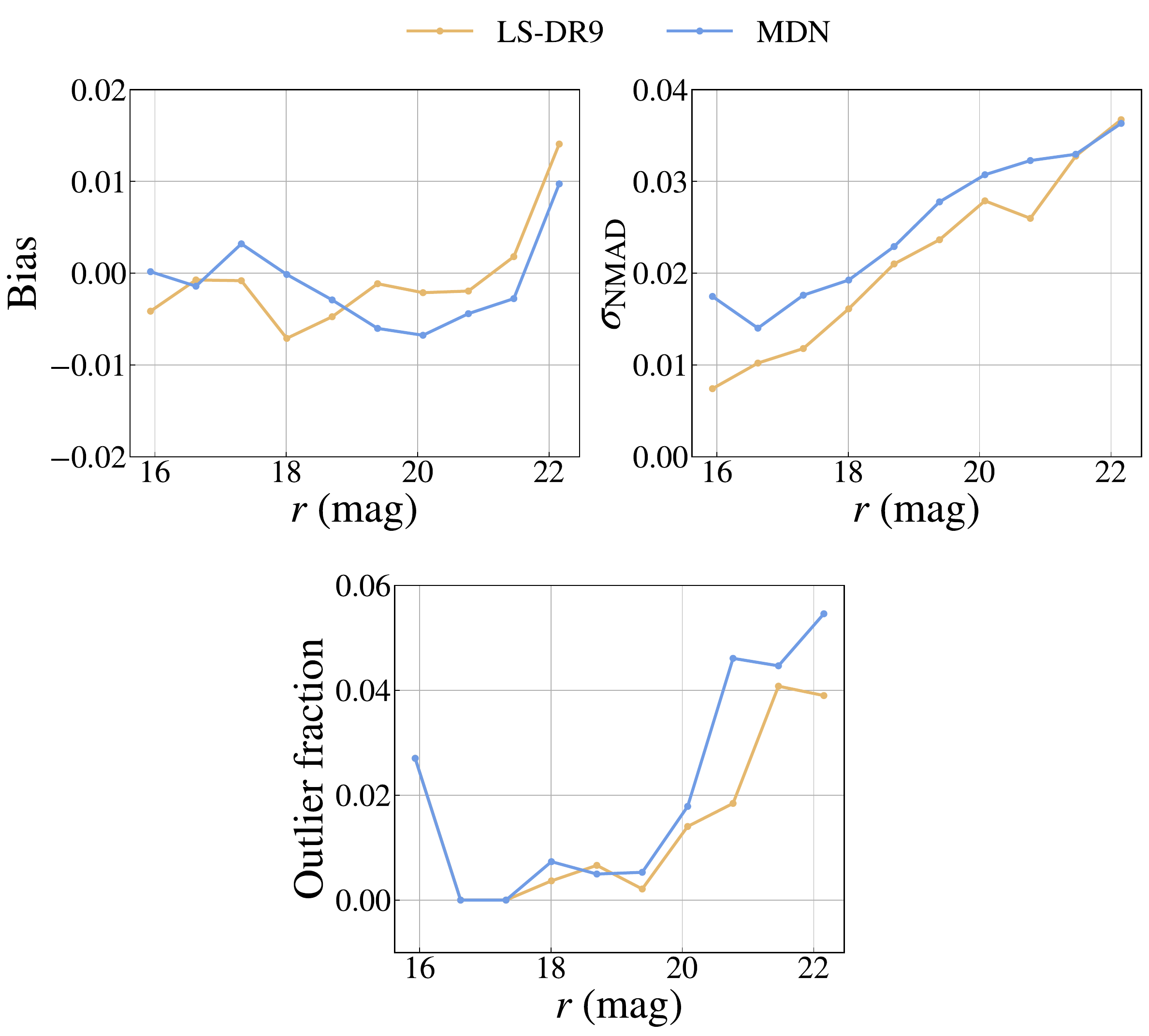}
    \caption{Performance of the photometric redshift estimates as a function of $r$-band magnitude. We compare the publicly available LS DR9 redshifts with those obtained using the MDN. The panels show the photo-$z$ bias, $\sigma_{\rm NMAD}$, and outlier fraction.}
    \label{fig:mdnmetrics}
\end{figure}

Figure~\ref{fig:mdnmetrics} presents the performance of the new photometric redshifts as a function of $r$-band magnitude, compared to the publicly available estimates. The new model achieves comparable performance in terms of the photo-$z$ bias (the averaged $\Delta z = z_{\rm photo} - z_{\rm spec}$), defined as the difference between the photometric redshift point estimate and the spectroscopic redshift of the same galaxy, and the outlier fraction (i.e., objects with $\Delta z / (1 + z_{\rm spec}) > 0.15$), with a slightly larger normalized median absolute deviation ($\sigma_{\rm NMAD} = 1.48 \times \mathrm{median}\left[ \left| \Delta z - \mathrm{median}(\Delta z) \right| / (1 + z_{\rm spec}) \right]$). 
Moreover, as shown in \citet{alfradique2024}, the use of full photo-$z$ PDFs increases the reliability of posterior estimates, making the new LS-DR9-north redshifts particularly well suited for this application. Further details on the development of these photometric redshifts, as well as the full catalogs, will be presented in \citet{Teixeira_inprep}.

\section{Method}
\label{sec:method}

The statistical dark siren method assumes that every bright galaxy within the GW localization volume is a potential host of the event, each assigned a probability of being the true host based on its position. By combining the redshift information from a galaxy catalog with the luminosity distance measurement inferred from the GW signal, one obtains an individual contribution to the posterior on $H_0$ from each possible host. The final $H_0$ constraint is then obtained by marginalizing over all galaxies in the localization volume. This approach was first proposed by \citet{schutz}, and \citet{delpozzo} was the first to apply it to simulated data. Later refinements of the method, including additional challenges associated with real data, are described in \citet{chen17}. In this work, we follow the statistical formulation presented in \citet{chen17}, with modifications introduced in later studies \citep{darksiren1, palmese20_sts, palmese2023, bom2024}. 

In the statistical dark siren scenario, we exploit the fact that GW data provide measurements of the luminosity distance, $\hat{d}_{L}$, and the GW source solid angle, $\hat{\Omega}_{\rm GW}$. In contrast, EM data provide measurements of galaxy redshift $\hat{z}_{i}$, solid angle $\hat{\Omega}_{i}$, and apparent magnitude $\hat{m}_{i}$ for each potential host galaxy \textit{i}. In this work, we denote measured quantities with a hat and their corresponding true values without.  

The joint likelihood of the electromagnetic observations from the galaxy catalog and the GW data can be expressed as the product of two individual likelihoods, $p\left(d_{\rm GW}, d_{\rm EM}| H_{0}\right) = p\left(d_{\rm GW}|H_0\right) p\left(d_{\rm EM}|H_0\right)$, since the measurements are independent, where $p(d_{\rm GW}\mid H_0)$ and $p(d_{\rm EM}\mid H_0)$ denote the GW and EM likelihoods, respectively. In the following, we construct this joint likelihood based on two assumptions: (i) the GW source is hosted by one of the galaxies in the catalog, and (ii) BBH mergers are expected to occur preferentially in more luminous galaxies, as luminosity in specific bands correlates with galaxy properties such as stellar mass and star formation rate. This second assumption is supported by results from synthetic simulations of compact binary mergers, which demonstrate a strong correlation between stellar mass, star formation rate, and the merger rate per galaxy in the local Universe \citep{Artale2019a, Artale2019b, Santoliquido2022, Rauf2023, Palfi2025}. The GW likelihood $p\left(d_{\rm GW}|H_0\right)$ is first marginalized over the true luminosity distance, sky location of the source, and other intrinsic parameters of the GW source (such as spins, masses, and inclination angle). We then marginalize the joint likelihood over the true redshifts, sky positions, and apparent magnitudes of all $N_{\rm gal}$ galaxies in the catalog, such that the joint likelihood can be written as \citep{chen17}:
\begin{eqnarray}
p(d_{\rm EM}, d_{\rm GW} \mid H_{0})
\propto
\frac{1}{\beta(H_{0})}
\sum_{i=1}^{N_{\rm gal}}
\int w_i(m_i,z_i) \nonumber\\
\quad \times
\,
p\!\left(d_{\rm GW}\mid d_L(z_i,H_0),\Omega_i\right)
\prod_{j=1}^{N_{\rm gal}}
p\left(d_{{\rm EM},j}|z_{j},\Omega_{j},m_{j} \right)  \nonumber\\
\quad \times
\frac{\psi(z_j)\, r^2(z_j,H_0)}
{(1+z_j)\, H(z_j,H_0)} dz_{j}d\Omega_{j}dm_{j}, 
\label{eq:joint_like}
\end{eqnarray}
where 
\begin{eqnarray}
    w_{i}\left(m_{i}, z_{i}\right) = \left| \frac{L_{i}}{L_{*}}\right| = 10^{-0.4\left(M_{i}\left(m_{i},z_{i}, H_0\right)-M_{*}\left(H_0\right)\right)}
\end{eqnarray}
denotes the statistical weight assigned to each galaxy representing the probability that it hosts the GW source, $L_{*}$ and $M_{*}$ are the “knee” luminosity and corresponding magnitude of the Schechter function, $\beta\left(H_0\right)$ is a normalization term that ensures the likelihood integrates to unity over all detectable GW and EM datasets, $r$ is the comoving distance, and $H\left(z, H_{0}\right)=H\left(z, H_{0}; \Omega^{\rm fid}_m=0.3\right)=H_{0}\left(0.3\left(1+z\right)^3+0.7\right)^{1/2}$ represents the Hubble parameter in a fiducial flat $\Lambda$CDM cosmology. The last term in eq.~(\ref{eq:joint_like}) represents the joint prior on the population-level parameters, defined as the product over all galaxies of their individual redshift priors. We assume that galaxies are uniformly distributed in comoving volume, with a merger rate evolution $\psi(z)$ following the Madau--Dickinson cosmic star formation rate \citep{Madau_2014}. The factor $(1+z_j)^{-1}$ accounts for the conversion from the source frame to the detector frame.

The statistical weight is based on the second assumption discussed above. As noted by \citet{Gray2023, gwtc4_cosmo}, although the transformation from absolute to apparent magnitude depends on cosmology through the relation $M_{i}\left(m_{i}, z_{i}, H_{0}\right) = m_{i} - 5\log_{10}\left(d_{L}\left(z_{i}, H_{0}\right)/ {\rm Mpc}\right) - 25 - K_{\rm corr}\left(z_{i}\right)$, the weights $w_{i}$ do not depend on $H_0$ since this dependence is eliminated in the ratio $L/L_{*}$, given that $L_{*}$ follows the same scaling with $H_0$. $K_{\rm corr}$ is the K-correction term, which corrects for the shift of a galaxy’s observed spectrum due to redshift, converting the observed magnitude into the equivalent rest-frame magnitude. This term was computed according to \citet{Kochanek2001}. We choose to use the galaxy \textit{r}-band luminosity for weighting, as it most closely corresponds to the near-infrared \textit{K}-band, which is believed to serve as a reliable proxy for stellar mass \citep{Singer_supp, Fong2013, Leibler2010}. 

Assuming that the observed EM quantities are conditionally independent given the true values, the EM likelihood for the \textit{j}-th galaxy can be written as
\begin{eqnarray}
    p\left(d_{{\rm EM},j}|z_{j},\Omega_{j},m_{j} \right) = p\left(\hat{\Omega}_j|\Omega_{j}\right) p\left(\hat{z}_j|z_{j}\right)p\left(\hat{m}_j|m_{j}\right).
    \label{eq:EMmargilike}
\end{eqnarray}
The galaxy positions are assumed to be perfectly measured and are represented by Dirac delta functions centered on the observed values. We model the likelihood of the apparent magnitude $m_j$ with a Gaussian distribution, $\mathcal{N}$, centered on the measured value $\hat{m}_j$ with standard deviation $\sigma_{m_j}$.

Following \citet{bayestar}, the marginal GW likelihood for the set of source parameters $\{z, \Omega\}$ can be approximated by a Gaussian function, expressed as
\begin{eqnarray}
p(d_{\rm GW}\mid d_L(z,H_0), \Omega_i)
\propto
\frac{p(\Omega_i)}{\sqrt{2\pi}\sigma_{d_{L}}(\Omega_i)}\nonumber\\
\times\exp\!\left[
-\frac{(d_L(z,H_0)-\hat{d}_{L}(\Omega_i))^2}{2\sigma_{d_{L}}^2(\Omega_i)}
\right],
\label{eq:gwlike}
\end{eqnarray}
where $p\left(\Omega_{i}\right)$ represents the sky position probability, and $\hat{d}_{L}\left(\Omega_{i}\right)$ and $\sigma_{d_{L}}\left(\Omega_{i}\right)$ correspond to the mean and standard deviation of the luminosity distance, respectively. The area of the \textit{i}-th HEALPix pixel on the sky is defined by $\Omega_{i}$.

The marginal GW likelihood as defined in eq. \ref{eq:gwlike} accounts only for a subset of the source parameters, $\vec{\theta}=\{z, \Omega\}$. This is based on the assumption that the remaining parameters of the GW waveform are sampled from a population distribution identical to the prior employed in the inference of individual detections. However, as detector sensitivity improves, leading to more precise measurements, and given our limited knowledge of the true population prior for the source-frame masses, this assumption may no longer hold. In such cases, it becomes necessary to explicitly model or marginalize over the possible discrepancies between the assumed and true mass distributions. To account for this, following \citet{Gray2023a, Gray2023, Borghi2024, gwtc4_cosmo} (see equation 18 of \citealt{Borghi2024}), we approximate the marginal GW likelihood (hereafter referred to as full) using a three-dimensional kernel density estimate (KDE), where the GW posterior, $p\left(z, m_{1}, m_{2}, \Omega|d_{\rm GW}\right)$, samples are re-weighted by the source-frame prior and the Jacobian of the source-to-detector frame transformation ($|d\vec{\theta}^{\rm det}\left(\vec{\theta},H_{0}\right)/d\vec{\theta}|$), as given by:
\begin{eqnarray}
    p\left(d_{\rm GW}|d_{L}\left(z,H_0\right), \Omega\right) = \int dm_{1} dm_{2} \frac{p\left(z, m_{1}, m_{2}, \Omega|d_{\rm GW}\right)}{p\left(m_{1}\right)p\left(m_{2}\right)p\left(d_{L}\right)}\nonumber\\
    \times\left(\frac{dd_{L}}{dz}\left(z, H_{0}\right)\right)^{-1}p\left(m_{1}, m_{2}|\vec{\lambda}_{m}\right),
    \label{eq:fullgwlike}
\end{eqnarray}
where the term $\frac{dd_{L}}{dz}\left(z, H_{0}\right) = \frac{d_{L}\left(z,H_0\right)}{1+z} + \frac{c\left(1+z\right)}{H\left(z,H_0\right)}$ is the Jacobian of the transformation from $z$ to $d_{L}$, and $p\left(m_{1}, m_{2}|\vec{\lambda}_{m}\right)$ is denotes the BBH mass population model characterized by the parameter set $\vec{\lambda}_{m}$. The priors and GW posteriors were taken from the results of \citet{gwtc4_inference}. In this work, we adopt the BBH Broken Power Law + 2 Peaks model \citep{Callister2024}, using the best-fit curve reported by the GWTC-4 population analysis \citep{gwtc4_pop}.

Substituting the definition of the marginal EM likelihood (eq. \ref{eq:EMmargilike}) into the joint likelihood (eq. \ref{eq:joint_like}) and factorizing the product, the joint likelihood can be rewritten and simplified, yielding the posterior for $H_0$ as:
\begin{eqnarray}
    p\left(H_{0}|d_{\rm GW}, d_{\rm EM}\right)  \propto \frac{p\left(H_0\right)}{\beta\left(H_0\right)}\sum_{i=1}^{N_{\rm gal}}\frac{1}{\mathcal{Z}_{i}} \int dz_{i} \int d\Delta z p\left(\Delta z\right)\nonumber\\
     \times p\left(d_{\rm GW}|d_{L}\left(z_{i}, H_{0}\right), \hat{\Omega}_{i}\right) p\left(\hat{z}_{i}|z_{i}, \Delta z\right)\frac{\psi\left(z_{i}\right)r^2\left(z_{i},H_{0}\right)}{\left(1+z_{i}\right)H\left( z_{i},H_{0}\right)}\nonumber\\
    \times 
    \left(\int_{m_{\rm min}}^{m_{\rm max}} dm_{i} 10^{-0.4\left(M_{i}\left(m_{i},z_{i}, H_0\right)-M_{*}\left(H_0\right)\right)} \mathcal{N}\left(\hat{m}_{i}|m_{i}\right)\right),
\end{eqnarray}
where 
\begin{eqnarray}
\mathcal{Z}_{i} =  \int d z_{i}\, p(\hat{z}_{i} \mid z_{i})
   \frac{\psi(z_{i})\, r^{2}(z_{i}, H_{0})}{(1+z_{i}) H(z_{i}, H_{0})}
\end{eqnarray}
is the evidence terms that correctly normalize the posterior, and $p\left(H_0\right)$ is the prior on $H_0$, which we assume to be uniform over the range [20, 140] km/s/Mpc. The integration over apparent magnitude is constrained by two limits: a bright-end cutoff defined by the \textit{r}-band Schechter function, $M_{\rm min}\left(z_{\rm min}, m_{\rm min}\right)$, evaluated at the minimum redshift of our galaxy sample, assuming a cosmology with the largest $H_0$ value from our prior; and a faint-end limit corresponding to the apparent magnitude threshold that characterizes the completeness limit of the galaxy catalog. For the LS, we adopt $m_{r, {\rm thr}} = 21$ for this threshold. As shown in Fig.~\ref{fig:schechter_function_Rband}, the \textit{r}-band absolute magnitude distribution of LS galaxies is well described by a Schechter function with parameter values consistent with those reported by \citet{Blanton2003}. We therefore adopt the corresponding \textit{r}-band Schechter parameters: $M_{*} = -20.44$ and $M_{\rm min} = -24.26$. In the equation above, the marginal EM likelihood for a given galaxy reduces to 
$p(\hat{m}_{i}\mid m_{i})\, p(\hat{z}_{i}\mid z_{i})
= \mathcal{N}(\hat{m}_{i}\mid m_{i})\, p(\hat{z}_{i}\mid z_{i})$, 
where $\mathcal{N}(\hat{m}_{i}\mid m_{i})$ denotes the Gaussian likelihood for the apparent magnitude measurement, and $p(\hat{z}_{i}\mid z_{i})$ represents the photometric redshift likelihood computed as described in Section~\ref{sec:photoz}. We also include a correction for the bias in photo-$z$, $\Delta z$, present in the data. This bias arises because the distribution of galaxies in redshift or color is not uniform beyond certain magnitude or color selection limits, causing the deep learning algorithm to preferentially sample redshifts near the peaks of the distribution and thus introducing systematic errors.

To account for selection effects in the dark siren analysis, we adopt the same definition presented in \citet{chen17, fishbach, mandel, Gray2020}. This effect is incorporated into a normalization factor that accounts for the biases introduced by the limited sensitivity of both the GW detectors and the survey telescopes, which allows them to observe only a fraction of the total population of compact binary coalescences and their potential host galaxies. Neglecting this correction could introduce significant biases in the inference of cosmological parameters, as discussed in \citet{mandel}. Under the assumption that GW events are uniformly distributed across the sky on cosmological scales, the normalization factor, $\beta\left(H_0\right)$, can be expressed in a compact form as (see \citep{chen17, mandel}):
\begin{equation}
    \label{fig:selectionfunction}
    \beta\left(H_0\right) = \int_{0}^{z_{\rm max}} p^{\rm GW}_{\rm sel}\left(d_{L}\left(z,H_{0}\right)\right)p^{\rm EM}_{\rm sel}\left(d_{L}\left(z,H_{0}\right)\right)p\left(z\right)dz,
\end{equation}
where $p\left(z\right)$ represents the prior redshift distribution of the galaxy catalog, $p^{\rm GW}_{\rm sel}$ represents the probability that a GW event at luminosity distance $d_{L}$ is detected by the network of detectors, and $p^{\rm EM}_{\rm sel}$ is the probability that a galaxy at $d_{L}$ is observed in the galaxy survey. We assume that the galaxy distribution is uniform in comoving volume, while accounting for the Madau–Dickinson phenomenological model for the star formation rate \citep{Madau_2014}. The dependence on sky position in the equation above is neglected under the assumption that the sources are isotropically distributed. As pointed out by \citet{chen17, palmese2023}, the galaxy distribution is not expected to be strongly correlated with the GW detector antenna pattern; hence, this assumption is not expected to introduce any significant systematic biases.

The term $p^{\rm GW}_{\rm sel}$ must include our selection criteria for the well-localized dark siren sample used in this analysis: $A_{50\%}<1000\,\deg^2$, $d_{L}<1500\,\rm{Mpc}$, and GW events with more than 70\% of their sky localization probability covered by the LS. However, because we do not expect the LS footprint to be correlated with the GW antenna pattern, this last condition does not need to be explicitly included in the computation. In practice, this is equivalent to performing an isotropic random sampling of the GW events. On the other hand, the term $p^{\rm EM}_{\rm sel}$ essentially describes the coverage of a given magnitude-limited catalog, as the EM likelihood is considered uninformative in the dark siren context, since no counterpart can be identified and associated with the GW source. The functions $p^{\rm GW}_{\rm sel}$ and $p^{\rm EM}_{\rm sel}$ are smooth and range from 0 to 1 as a function of redshift, for different assumed values of $H_0$.  

\subsection{Gravitational wave and electromagnetic selection functions}\label{sec:emselfunc}

To model the selection function, we adopted an approach consistent with previous dark siren analyses \citep{bom2024, alfradique2024, palmese2023}. In our implementation, we generated a synthetic catalog of 100,000 BBH merger events using the \texttt{BAYESTAR} software \citep{bayestar, Singer_2016, Singer_supp}, employing the IMRPhenomD waveform model in the frequency domain. The BBH population was drawn from a Broken Power Law + Two Peaks mass distribution, using the best-fit curve reported in \citet{gwtc4_pop}, and assuming uniform spin distributions in the range $\left(-1, 1\right)$. The BBH population was distributed uniformly in comoving volume, assuming the Madau–Dickinson parametrization to describe the redshift evolution of the merger rate. As in the O4a observing run, Virgo and KAGRA were in the commissioning phase (with KAGRA operating for only one month); therefore, the matched-filter analysis used to compute the signal-to-noise ratio (SNR) for each injected event was performed using only the LIGO Hanford and Livingston detectors. The analysis employed the power spectral density estimates provided in \citet{psdsO4a} and described in \citet{gwtc4_psd}. We adopted a detection criterion requiring a network SNR greater than 12 and an individual detector SNR greater than 4. To mimic realistic measurements, Gaussian noise was added to all injected GW signals. For each detected event, we used \texttt{BAYESTAR} to reconstruct the sky map, assuming a luminosity distance prior proportional to $d_{L}^{2}$. We performed 25 independent GW simulations, each assuming a different $H_0$ value within our prior range. The resulting function was interpolated to guarantee a continuous definition over the entire $H_0$ range.

\begin{figure}
    \centering
    \includegraphics[width=\linewidth]{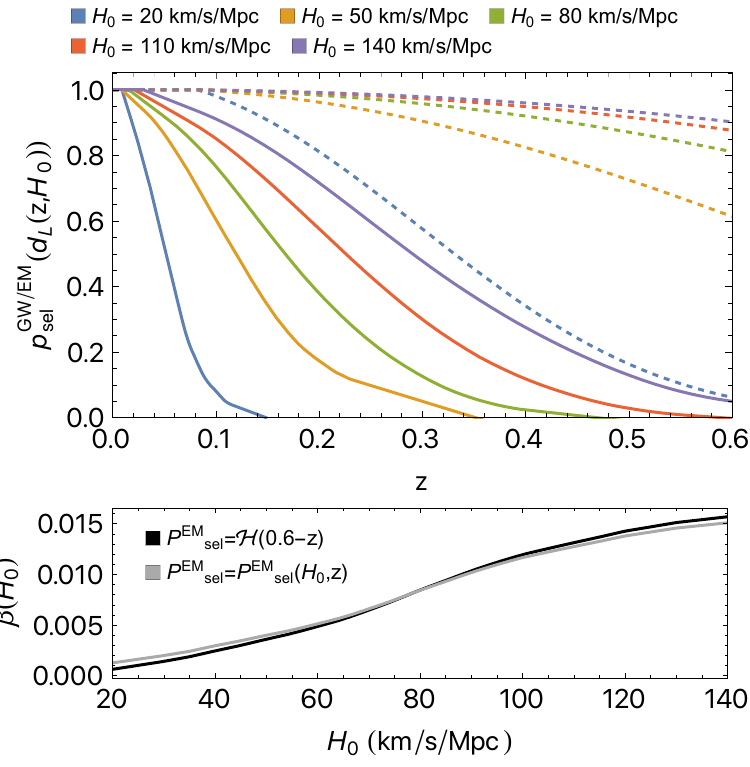}
    \caption{GW and EM selection functions. \textbf{Top panel}: gravitational wave (solid lines) and electromagnetic (dashed lines) detection probabilities as functions of $H_0$ (shown in different colors) computed for an apparent magnitude threshold of $m_{r, \rm{thr}}=21$. \textbf{Bottom panel}: Selection function (as defined in eq. \ref{fig:selectionfunction}) computed from GW detection simulations using \texttt{BAYESTAR}. The black curve shows the result including only the GW detection probability (solid lines in the top panel), while the gray curve includes the EM detection probability as well (dashed lines in the top panel).}
    \label{fig:pdet_EM}
\end{figure}

In the dark siren methodology, the electromagnetic data correspond to the identification of potential host galaxies, in contrast to the bright siren case, where the electromagnetic counterpart arises from the transient flare produced at the time of the merger, enabling a direct host identification. In this context, the term $p_{\rm sel}^{\rm EM}$ accounts for the incompleteness of the magnitude-limited galaxy catalog, in contrast to the bright siren case, where it instead represents the probability of detecting an electromagnetic counterpart and must be modeled accordingly to avoid potential systematics related to EM selection effects and their possible mitigation strategies (see discussions in \citep{Chen2020, Mancarella2024, Salvarese2024, Muller2024, Chen2024}). Here, we compute the galaxy catalog incompleteness for the LS following the procedure of \citet{gwtc4_cosmo}. As established in the previous subsection, the \textit{r}-band absolute magnitude distribution is well described by a Schechter function with parameter values derived from the SDSS catalog (see Fig.~\ref{fig:schechter_function_Rband}). We then adopt the incompleteness definition given in Appendix B2 of \citet{gwtc4_cosmo} (their eq.~B20) to evaluate the completeness fraction as a function of $z$ and $H_0$ for an apparent magnitude threshold of $m_{r,\mathrm{thr}} = 21$. The resulting completeness curves are shown as dashed lines in the top panel of Fig.~\ref{fig:pdet_EM}. In the bottom panel, we present two versions of the selection function: (1) considering only the GW detection probability and assuming a Heaviside step function for $p_{\rm sel}^{\rm EM}$ at the redshifts of interest (black line), and (2) including the incompleteness of the magnitude-limited galaxy catalog in $p_{\rm sel}^{\rm EM}$ (gray line). The two curves differ only slightly, with the relative difference remaining at the percent level across the entire $H_0$ prior range.

\subsection{Galaxy fakes}{\label{sec:fake}}

The lack of galaxies in the catalog, caused by the sensitivity limitations of optical telescopes, can introduce incompleteness effects, which typically bias the sample toward the bright end of the luminosity distribution and lead to a bias toward a lower $H_0$ value (as discussed in \citealt{Alfradique2025}). To correct this effect, it was suggested that the missing galaxies be assumed to be uniformly distributed in comoving volume \citep{chen17, fishbach, Gray2020, LVC_O2_StS, Abbott_2023, Mastrogiovanni2023, Borghi2024, posteriorO4a}. However, other studies have noted that assuming uniform completeness introduces biases in the $H_0$ estimation in the regime where the out-of-catalog term dominates, and have proposed extensions that account for inhomogeneities in the galaxy distribution along the line of sight \citep{Finke2021, Gray2023a}. Following this approach, \citet{Leyde2025} recently proposed a method that reconstructs the missing galaxy distribution by assuming that this field follows the two-point statistics. As shown by \citet{Alfradique2025} using simulated detections for the O4 and O5 observing runs, it is possible to obtain an unbiased estimate of $H_0$ with a volume-limited catalog. Based on this result, and following the approach adopted in previous works (and described in Section \ref{sec:desi}), we construct our galaxy catalog to be volume-limited and therefore do not include an additional correction term in the dark siren methodology. Additionally, we inject synthetic galaxies into the regions not covered by the photo-\textit{z} catalogs to mitigate residual incompleteness. Among our dark siren sample, three GW events (GW230919\_215307, GW230927\_153832, and GW230731\_215307) have 90\% CI areas that approach our selection threshold of 70\% coverage by the LS catalog.

To mitigate the impact of these uncovered regions, we adopt the procedure described in \citet{palmese2023, bom2024} to inject synthetic galaxies into the missing areas. The synthetic galaxies are assigned redshifts sampled from the prior defined by the training sample using a Monte Carlo procedure, while their positions are uniformly distributed across the sky with a number density set by the mean density of the LS galaxy catalog. To assign uncertainties in photo-$z$ and apparent magnitude, we first use our training sample to compute the mean and standard deviation within photo-$z$ bins of width 0.05, and then derive the relation between photo-$z$ error or apparent magnitude and photo-$z$. We assume that the photo-$z$ PDFs of the synthetic galaxies are Gaussian, sampling each value around its assigned true redshift with a standard deviation set by the corresponding photo-$z$ uncertainty. In section \ref{sec:results}, we discuss the impact of these synthetic galaxies on our final $H_0$ constraint.
  
\section{Results and Discussion} \label{sec:results}

In this section, we report our estimates of the Hubble constant obtained from the seven O4a gravitational wave events using the statistical dark siren framework introduced in Section~\ref{sec:method}. The analysis combines GW data with photometric redshift probability distributions from the LS, constructed with the deep-learning approach developed in our previous work \citep{alfradique2024, bom2024, teixeira2024} and described in Section~\ref{sec:photoz}. Our results are presented both for individual events and for their joint combination. We focus primarily on two methodological effects in the analysis: (i) the inclusion of $r$-band luminosity weighting, and (ii) two different implementations of the GW likelihood. These are: (1) a Gaussian approximation that depends only on the source parameters $\{d_{L}, \Omega\}$, as defined in eq.~\eqref{eq:gwlike}; and (2) the full likelihood obtained from a KDE of the posterior samples, reweighted by the population prior on the source-frame masses, as in eq.~\eqref{eq:fullgwlike}. In addition, we explore the impact of electromagnetic selection effects and of using photo-$z$ PDFs obtained with the MDN compared to a Gaussian photo-$z$ approximation (see the discussion in App.\ref{app:lum_eff}).

\begin{figure*}
    \centering
    \includegraphics[width=\linewidth]{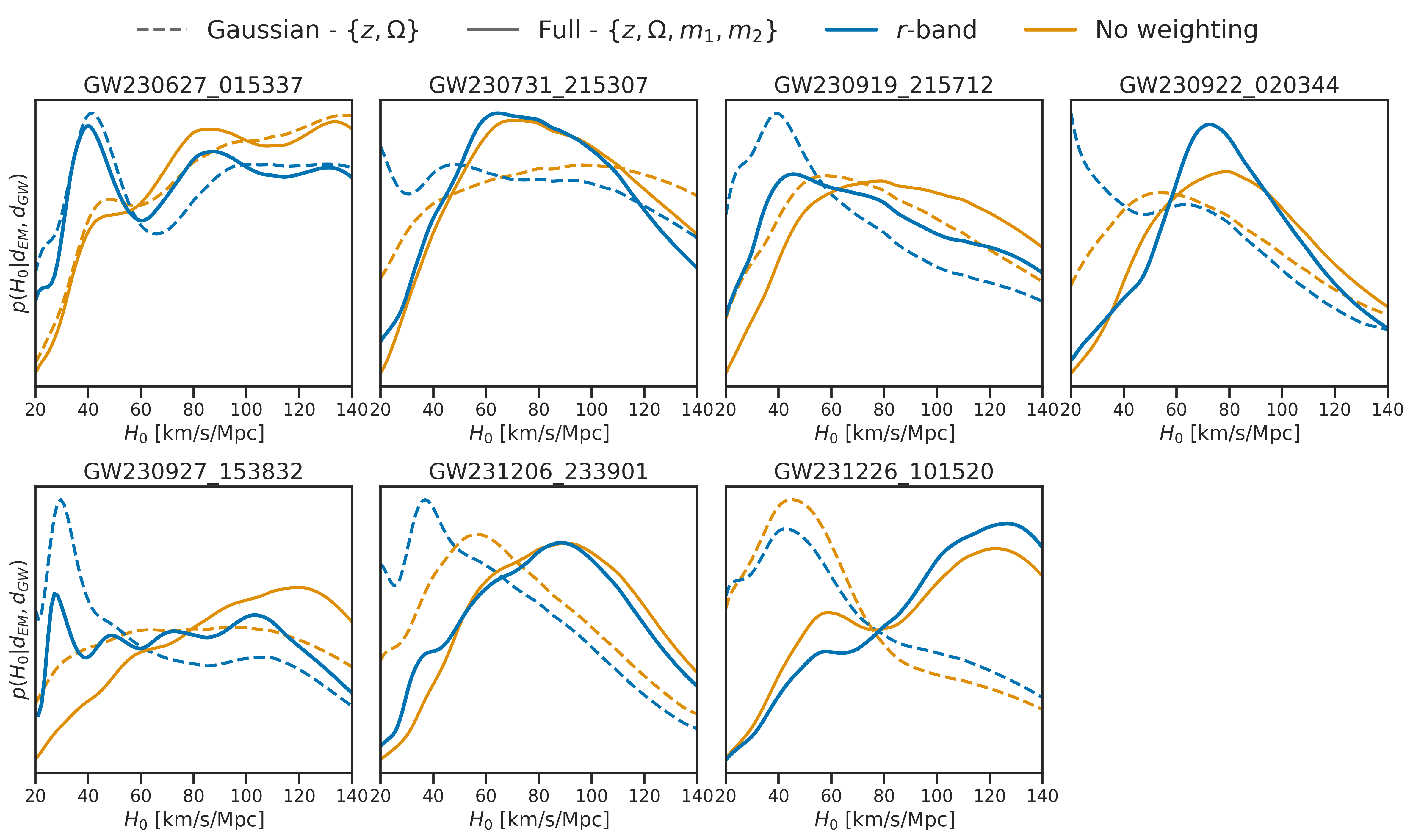}\\
    \caption{Posterior distributions of the Hubble constant inferred from individual gravitational wave events. Solid lines represent the analysis using the full GW likelihood, and dashed lines represent the Gaussian approximation. Blue curves correspond to \textit{r}-band weighted results, while orange curves are unweighted.}
    \label{fig:pH0_all}
\end{figure*}

Figure~\ref{fig:pH0_all} shows the individual $H_0$ posteriors for the seven O4a GW events analyzed in this work. The most informative event is GW231206\_233901, and its 1$\sigma$ constraint reaches close to 77\% (74\%) of the $H_0$ prior in the unweighted ($r$-band weighted) case using the full GW likelihood. In contrast, GW230731\_215307 and GW230919\_215712 yield the weakest constraints: GW230731\_215307 leaves 85\%(82\%) of the prior width, and GW230919\_215712 leaves 89\%(91\%) in the same configurations. These events are also among the worst localized in the sample and suffer from limited galaxy-catalog coverage, which plausibly contributes to their broader $H_0$ posteriors. The peaks seen in the individual $H_0$ posteriors reflect overdensities in the redshift distribution of potential host galaxies along each GW line of sight, and the width of each peak depends on the precision of the galaxy photo-$z$ PDFs and on the size of the GW localization volume, which sets the number of potential host galaxies over which the marginalization is performed.

We quantify the impact of adding fake galaxies for the three GW events that require them (GW230731\_215307, GW230919\_215712, and GW230927\_153832). For individual events, the inclusion of fake galaxies flattens the $H_0$ posteriors, increasing the uncertainty by approximately $4\%$–$9\%$. When these events are combined with the full dark siren sample, the effect on the $H_0$ constraint becomes much smaller: the posterior mode shifts by at most $\sim 2\%$, and the uncertainties increase by no more than $\sim 1\%$ across all configurations explored in our analysis. No significant change is found when the results are combined with the bright siren GW170817.

We find statistical agreement with our previous results using the same configuration (no luminosity weighting and a Gaussian GW likelihood), when the five dark sirens had not yet been confirmed as GW events. In this analysis, the photo-\textit{z} model for the two GW events in the northern hemisphere (GW230627\_015337 and GW231226\_101520) was trained using only galaxies from the northern footprint of the LS DR9, whereas the previous work used a combined north and south DR9 training sample. This change alters the redshift distribution of potential host galaxies, shifting the mean redshift by $\Delta z_{\rm mean}=-0.037$ for GW230627 and $\Delta z_{\rm mean}=0.046$ for GW231226. Nevertheless, the resulting $H_0$ posterior remains consistent with our previous results within statistical uncertainties. The updated sky maps yield modestly improved 90\% CI areas, while in most cases the luminosity distance uncertainties are slightly larger. We found that these updates lead to tighter $H_0$ uncertainties, with relative reductions ranging from approximately 1\% to 23\% for the individual events, indicating that improvements in sky localization tend to have a larger impact than variations in the luminosity distance uncertainty.

\begin{table*}
\caption{Hubble constant constraints obtained from GW standard sirens, presented in this work and in the literature. All $H_0$ measurements and uncertainties are reported in km/s/Mpc. Reported uncertainties represent the 68\% highest-density interval around the posterior maximum. The value shown in parentheses indicates the relative precision, defined as $\sigma_{H_0}/H_0$. For dark siren analyses employing population modeling, ``PLP'' denotes the \textit{Power-Law + Peak} mass model \citep{Talbot2018, Fishbach2018}, and the semiparametric B-spline mass model corresponds to that introduced by \citet{Edelman2022, Edelman2023}. The ``BPL+3peak'' model is based on the ``FullPop-4.0'' compact binary population model \citep{gwtc4_pop,gwtc4_cosmo}, in which the broken power-law mass spectrum is supplemented with three Gaussian components (\citealt{Pierra2026}, hereafter BPL+3peak). References: 
(1) \citet{Abbott_2023}, 
(2) \citet{gwtc4_cosmo}, 
(3) \citet{Pierra2026}, 
(4) \citet{Tagliazucchi2026}, 
(5) \citet{bom2024}, 
(6) adapted from \citet{nicolaou2019impact}, 
(7) \citet{Mukherjee_2021}, 
and (8) this work.}
\centering
\resizebox{\textwidth}{!}{
\begin{tabular}{cccc}
\hline\hline
{\bf\Large Event(s)} & {\bf\Large Method(s)} & {\bf\Large $H_0\,(\rm{km/s/Mpc})$} & {\bf\Large $\sigma_{H_0}\,(\rm{km/s/Mpc})$} \\
\hline \hline
O1-O3 -- 47 dark sirens $^1$ & Catalog (K-band luminosity-weighted) + BNS/NSBH/BBH population (PLP model with fixed parameters)  & $67^{+13}_{-12}$ & $12.5$ $(18\%)$ \\
O1-O4a -- 142 dark sirens $^2$ & Catalog (K-band luminosity-weighted) + BNS/NSBH/BBH population (FullPop-4.0 model) & $81.6^{+21.5}_{-15.9}$ & $18.7$ $(23\%)$ \\
O1-O4a -- 142 dark sirens $^3$ & Catalog (K-band luminosity-weighted) + BNS/NSBH/BBH population (BPL+3peak model) & $82.5^{+16.8}_{-14.3}$ & $15.6$ $(19\%)$ \\
O1-O4a -- 142 dark sirens $^2$ & BNS/NSBH/BBH population (FullPop-4.0 model) & $76.4^{+23.0}_{-18.1}$ & $20.6$ $(27\%)$ \\
O1-O4a -- 137 dark sirens $^4$ & BBH population (semiparametric B-spline model) & $57.8^{+21.9}_{-20.6}$ & $21.3$ $(37\%)$ \\
O1-O4a - 15 dark sirens $^5$ & Catalog (no luminosity weighting) & $70.4^{+13.6}_{-11.7}$ &     $12.6$ $(18\%)$  \\
GW170817 $^6$ & Bright ($v_p$ corrected) & $68.8 ^{+ 17.3}_{-7.6}$ & $12.5(18\%)$  \\
GW170817 $^7$ & Bright ($v_p$+VLBI) & $68.3 ^{+4.6}_{-4.5}$ & $4.6(7\%)$ \\
\hline \hline
O1-O4a -- 17 dark sirens $^8$ & Catalog (no luminosity weighting; GW likelihood Gaussian approximation, eq.~\ref{eq:gwlike})  & $72.6^{+11.6}_{-11.3}$ & $11.4$ $(16\%)$ \\
O1-O4a -- 17 dark sirens $^8$ & Catalog ($r$-band luminosity-weighted; GW likelihood Gaussian approximation, eq.~\ref{eq:gwlike}) & $72.8^{+12.6}_{-14.3}$ & $13.5$ $(18\%)$ \\
O1-O4a -- 17 dark sirens $^8$ & Catalog (no luminosity weighting; full GW likelihood, eq.~\ref{eq:fullgwlike}) & $78.8^{+14.6}_{-12.2}$ & $13.4$ $(17\%)$ \\
O1-O4a -- 17 dark sirens $^8$ & Catalog ($r$-band luminosity-weighted; full GW likelihood, eq.~\ref{eq:fullgwlike}) & $78.2^{+12.0}_{-11.0}$ & $11.5$ $(15\%)$ \\
O1-O4a -- 17 dark + 1 bright siren $^8$ & Bright (host) + Catalog ($r$-band luminosity-weighted; full GW likelihood, eq.~\ref{eq:fullgwlike}) & $75.2^{+8.0}_{-7.9}$ & $8.0 (11\%) $  \\
O1-O4a -- 17 dark + 1 bright siren (EM) $^8$ & Bright ($v_p$+VLBI) + Catalog ($r$-band luminosity-weighted; full GW likelihood, eq.~\ref{eq:fullgwlike}) & $69.9^{+4.1}_{-4.0}$ & $4.1$ $(6\%) $ \\ 
\hline \hline
\end{tabular}}
\label{tab:results}
\end{table*}

\begin{figure*}
    \centering
    \includegraphics[width=\linewidth]{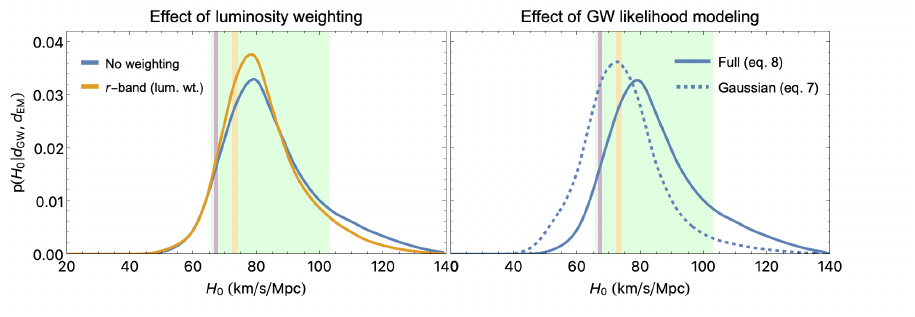}
    \caption{\textbf{Left panel:} effect of galaxy-catalog weighting, with the full GW likelihood held fixed. The blue curve shows the result with no weighting, the orange curve uses $r$-band luminosity weighting, and the green curve uses $g$-band luminosity weighting. \textbf{Right panel:} effect of the GW likelihood treatment using the unweighted catalog. The solid blue curve corresponds to the full GW likelihood (as defined in eq.~\ref{eq:fullgwlike}), while the dashed blue curve shows the Gaussian approximation likelihood (see eq.~\ref{eq:gwlike}). In both panels, shaded regions indicate the 68\% CI on $H_0$ from the Planck CMB inference \citep{planck18} (pink), the SH0ES distance-ladder measurement \citep{Breuval2024} (yellow), and the GWTC-4 dark siren analysis based on 142 events with the FullPop-4.0 population model \citep{gwtc4_cosmo} (green).}
    \label{fig:lum_gwlike}
\end{figure*}

\begin{figure*}
    \centering
    \includegraphics[width=0.8\linewidth]{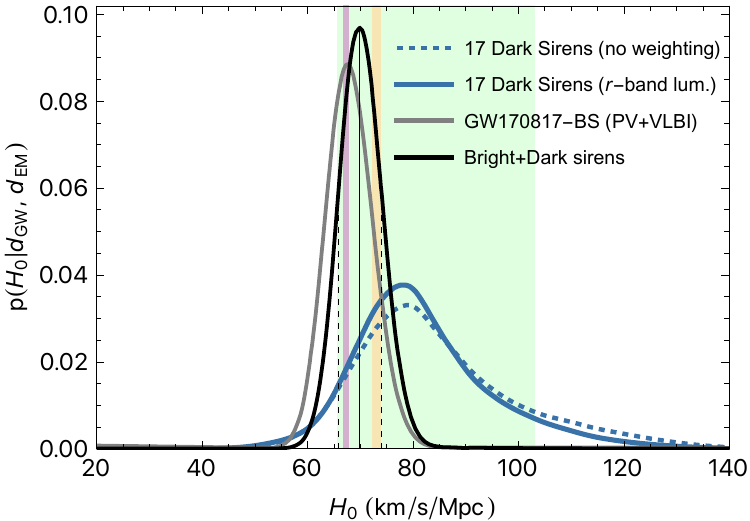}
    \caption{Hubble constant posterior inferred from dark and bright GW standard sirens. The blue curves show the constraints from the 17 dark sirens observed during O1–O4a and analyzed in this work using the full GW likelihood: the solid blue line corresponds to the result obtained with $r$-band luminosity weighting of the galaxy catalog, while the dashed blue line assumes equal weights. The gray curve shows the bright siren measurement from GW170817 using the GW+VLBI data with peculiar velocity corrections \citep{Mukherjee_2021}. The black curve represents the joint constraint from combining the bright siren with the luminosity-weighted dark siren sample. For reference, shaded regions indicate the $1\sigma$ intervals from the Planck CMB inference \citep{planck18} (pink), the SH0ES distance-ladder measurement \citep{Breuval2024} (yellow), and the GWTC-4 dark siren analysis based on 142 events with the FullPop-4.0 population model \citep{gwtc4_cosmo} (green).}
    \label{fig:result}
\end{figure*}

\subsection{Luminosity weighting effect}

We now investigate the impact of \textit{r}-band luminosity weighting on the individual $H_0$ posteriors and their combined constraint. Uncertainties in the apparent magnitudes of host galaxies have a negligible impact on the inferred $H_0$ posterior. The relative difference is below the $0.01\%$ level, and the apparent magnitude integral computed assuming perfect measurements differs from the result including magnitude uncertainties by at most $\mathcal{O}(10^{-6})$. The two results are therefore numerically indistinguishable, indicating that neglecting apparent magnitude errors does not affect the subsequent $H_0$ inference. 

For the events GW230731\_215307, GW230919\_215307, and GW230927\_153832, synthetic galaxies are injected as described in Section~\ref{sec:fake}. Since our deep learning model estimates only photo-\textit{z}, their apparent magnitudes are assigned through an empirical $m(z)$ relation derived from the training sample rather than a joint luminosity–redshift distribution. We find that applying luminosity weighting systematically shifts the host redshift distribution toward higher $z$. To avoid introducing this potential bias in the $H_0$ inference, we adopt a conservative approach and assign equal weights to the synthetic galaxies when computing the final $H_0$ posterior.

Overall, luminosity weighting modifies the redshift distribution of potential host galaxies by shifting the peak toward slightly higher redshift, with typical displacements of $\Delta z \sim 0.01$–$0.1$, and by making the distribution more concentrated (i.e., with a reduced effective width), as illustrated by the comparison between the blue and orange curves in Fig.~\ref{fig:pem_weight_int}. These changes propagate into the inferred $H_0$ posterior. The overall structure of the posterior is largely preserved, with luminosity weighting primarily reweighting existing modes rather than significantly modifying its shape (see the comparison between the orange and blue curves in Fig.~\ref{fig:pH0_all}). Quantitatively, the Kullback--Leibler (KL) divergence\footnote{The KL divergence \citep{Kullback1951, CoverThomas1991} between two probability densities $p(x)$ and $q(x)$ is defined as $D_{\mathrm{KL}}(P\|Q)=\int p(x)\,\log\!\big[p(x)/q(x)\big]\,dx$.} between the weighted and unweighted posteriors ranges between $D_{\rm KL}\simeq0.005$ and $0.39$ across the analyzed events, indicating moderate redistributions of probability density without qualitatively altering the posterior structure.

In several events, luminosity weighting shifts the $H_0$ posterior toward lower values, despite shifting the redshift distribution toward higher $z$. This behavior is observed, for example, in GW230627\_015337, GW230922\_020344, and GW231206\_233901. Nevertheless, the overall impact of luminosity weighting remains subdominant relative to other elements of the methodology (see the discussion in App.~\ref{app:lum_eff}).

In the luminosity-weighted analysis, combining the seven dark sirens from the O4a observing run with the ten events from previous runs and studies \citep{darksiren1, palmese20_sts, palmese2023, alfradique2024} yields $H_0 = 78.2^{+12.0}_{-11.0}\ \mathrm{km/s/Mpc}$ when using the full GW likelihood (see blue solid line in Fig.~\ref{fig:result}), and $H_0 = 72.8^{+12.6}_{-14.3}\ \mathrm{km/s/Mpc}$ when the GW likelihood is approximated by a Gaussian. These measurements are consistent within $1\sigma$ with each other and independent determinations of $H_0$, including those from SH0ES, Planck, and the GWTC-4 dark siren analysis. In Fig.~\ref{fig:lum_gwlike} we compare the luminosity-weighted and unweighted $H_0$ posteriors. When using the full GW likelihood, the inclusion of luminosity weighting reduces the uncertainty on $H_0$ by 1.9 km/s/Mpc, corresponding to an improvement of $\sim14\%$. 

Figure \ref{fig:result} presents our main results (summarized in Table~\ref{tab:results}), shown as blue solid and dashed lines for the 17 dark sirens analyzed with the full GW likelihood. The combination of the bright siren GW170817 measurement (gray line), which accounts for peculiar velocity corrections and includes the inclination constraint from VLBI data \citep{Mukherjee_2021}, with the luminosity-weighted dark siren posterior yields $H_0 = 69.9^{+4.1}_{-4.0}\,\rm{km/s/Mpc}$, and is shown as the black solid line in Fig.~\ref{fig:result}. The dark siren information provides a slightly broader posterior, with an improvement of 11\% in the uncertainty compared to the bright siren alone. Our results are broadly consistent within $1\sigma$ with those reported by the Planck \citep{planck18} and SH0ES collaboration \citep{Breuval2024} measurements. 

In addition, we compare our results with previous dark siren analyses based on the latest GW catalog release \citep{gwtc4_inference,Pierra2026,Tagliazucchi2026}. Here, we focus the comparison on our full GW likelihood result, since it is more directly comparable to previous analyses that also marginalize over the BBH mass population. Our result, based on the full GW likelihood, is broadly consistent with those studies, but we obtain a competitive and precise constraint despite the smaller number of events considered (see Table~\ref{tab:results}). The larger difference between our uncertainties and those reported in \citet{gwtc4_cosmo} arises primarily because they infer $H_0$ jointly with the BBH mass-population parameters, whereas in our analysis these parameters are fixed. Fixing the population parameters leads to more optimistic constraints, as also seen in the GWTC-3 results. The impact of this coupling additionally depends on the choice of BBH mass model. \citet{Pierra2026}, who adopt the same inference framework as \citet{gwtc4_cosmo} but use a better-fitting mass model (BPL+3peak, with a Bayes factor of 6 over the FullPop-4.0 model used in \citep{gwtc4_cosmo} when reweighted by the effective prior volume), report an improvement in the $H_0$ uncertainty of 36.2\% relative to \citet{gwtc4_cosmo}, yielding an uncertainty comparable to that obtained in our analysis.

In previous studies \citep{gwtc4_cosmo, Pierra2026, Abbott_2023}, the contribution from galaxy information has been found to be small compared to that from the BBH mass distribution, leading to improvements of only $\sim$7–9\% in the $H_0$ uncertainty. This lack of information can be understood by the choice of galaxy catalogs employed: those analyses relied on the GLADE+ catalog \citep{gladeplus}, which becomes highly incomplete and inhomogeneous at the distances typical of GWTC-4 events, well beyond its completeness limit of $\sim 47^{+4}_{-2}$ Mpc, leaving most events with insufficient support in the redshift range of interest. Therefore, following previous works \citep{palmese2023, alfradique2024, bom2024}, we restrict our analysis to well-localized events that lie within the completeness range of the DESI imaging survey, thereby avoiding the lack of redshift support that would otherwise yield a weakly constrained $H_0$ posterior.

\subsection{Effect of the BBH mass posteriors on the GW likelihood}

In this subsection, we investigate how different implementations of the GW likelihood impact the dark siren inference of $H_0$. For almost all GW events considered, we find that the full likelihood consistently shifts the peak of the $H_0$ posterior toward higher values compared to the Gaussian approximation, highlighting the impact of the two different GW likelihood implementations (see Fig.~\ref{fig:pH0_all}, where the unweighted case is shown in orange, and the dashed and solid curves correspond to the Gaussian and full likelihoods, respectively). Combining the 17 dark siren events and assuming equal probability for all possible host galaxies, we obtain $H_0 = 72.6^{+11.6}_{-11.3}\ \mathrm{km/s/Mpc}$ using the Gaussian likelihood, while the full likelihood yields
$H_0 = 78.8^{+14.6}_{-12.2}\ \mathrm{km/s/Mpc}$. 
These results are statistically consistent with each other within their respective uncertainties and are shown in the right panel of Fig.~\ref{fig:lum_gwlike}, indicating that the Gaussian approximation to the GW likelihood can still yield a consistent $H_0$ inference.

\begin{figure}
    \centering
    \includegraphics[width=\linewidth]{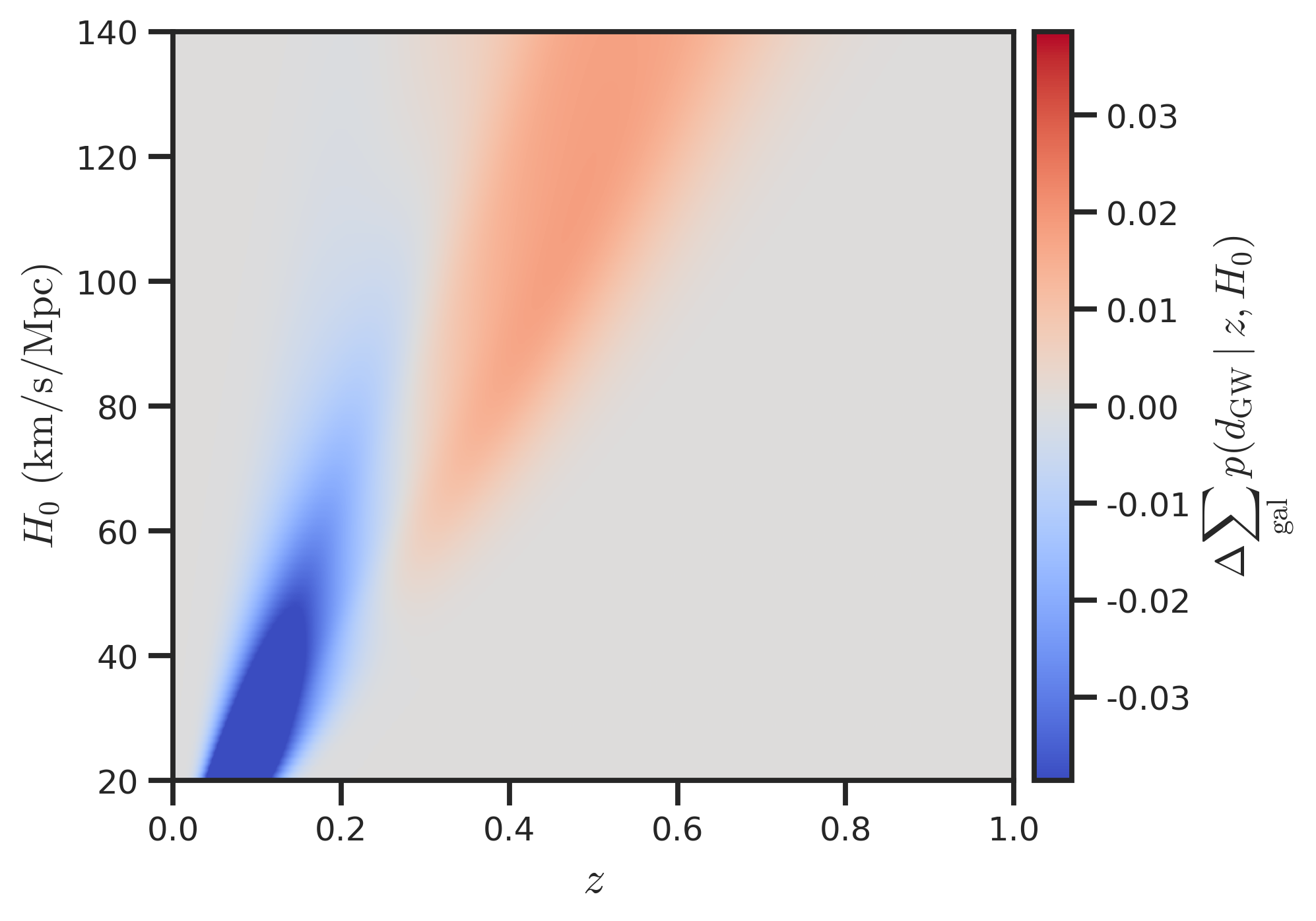}
    \caption{Difference in the galaxy-summed GW likelihood,
$\Delta \sum_{N_{\rm gal}} p(d_{\rm GW} \mid z, H_0)$,
between analyses using the BBH broken power-law + two-peaks mass model and assuming a flat source-frame mass prior for the event GW230922\_020344. Both likelihood surfaces are normalized over the full $(z,H_0)$ plane before differencing, so the map highlights only changes in the shape of the GW likelihood induced by the mass model. Red (blue) regions indicate where the mass model increases (decreases) the GW likelihood.}
    \label{fig:massprioreffect}
\end{figure}

The behavior seen in Fig.~\ref{fig:massprioreffect} is influenced by the inclusion of the BBH population mass prior in the GW likelihood. The mass model adopted in this work follows a broken power-law with two peaks for the primary black hole mass, with characteristic scales at $\sim 10\,M_\odot$ and $\sim 35\,M_\odot$, and therefore favors source-frame primary masses near these values. For any GW event, only specific combinations of $z$ and $H_0$ map the inferred source-frame masses, $m_i = m_{i,\rm det}/(1+z)$, into the preferred regions of the population mass distribution. 

Figure~\ref{fig:massprioreffect} illustrates this effect for GW230922\_020344 by comparing the GW likelihood obtained using the BBH broken power-law + two-peaks mass model to that obtained assuming a flat mass prior. The inclusion of the mass prior redistributes probability in the $(z,H_0)$ plane, enhancing regions where the inferred source-frame masses align with the preferred population scales and suppressing others. This demonstrates how the assumed mass model directly shapes the $(z,H_0)$ dependence of the GW likelihood and, consequently, the inferred $H_0$ posterior. As a result, likelihood support is reduced at low $H_0$ and low $z$ (blue regions) and enhanced at higher $H_0$ and redshift (red regions), shifting the inferred $H_0$ posterior toward higher values.

\subsection{Electromagnetic selection effects in dark siren analysis}

To quantify the impact of electromagnetic selection on the dark siren $H_0$ inference, we compare two selection functions, as described in Section~\ref{sec:emselfunc}: (1) one that includes only the GW detection cuts, assuming all galaxies are observable within the survey sensitivity at the distances of the dark siren events considered in this work, and (2) one that incorporates both the GW cuts and the incompleteness of the magnitude-limited galaxy catalog. The comparison of individual $H_0$ posteriors computed using the two selection functions reveals only a percent-level shift in the posterior mode, with uncertainties remaining $\lesssim 3\%$ for all dark siren events and configurations explored. We further quantify the difference between the posteriors using the KL divergence, finding $D_{\mathrm{KL}} < 3\times10^{-3}$, which indicates that the two distributions are nearly indistinguishable in both shape and information content. These small differences indicate that assuming all galaxies are observable within the survey sensitivity is a good approximation, yielding $H_0$ constraints that are fully consistent with each other and well within the statistical uncertainty.

\section{Conclusions}\label{conclusions}
In this work, we present a revised measurement of the Hubble constant using the statistical dark siren method, combining the best localized events from the O4a observing run of the LVK network in the GWTC-4 catalog with galaxy photometric data from the LS. Our analysis focuses on several key methodological aspects: in the electromagnetic component, the treatment of galaxy photometric redshifts and luminosity weighting, and in the GW component, alternative implementations of the GW likelihood. We investigate how these choices affect the inferred $H_0$ posteriors for both individual events and their combination. 

Using a sample of 17 dark siren events, seven from the O4a observing run and the other ten from previous runs analyzed in \citet{darksiren1, palmese20_sts, alfradique2024, bom2024}, we obtain a constraint of $H_0 = 78.8^{+14.6}_{-12.2}\,\rm{km/s/Mpc}$ corresponding to a precision of approximately $17\%$, when using the full GW posterior information and assuming no luminosity weighting for the potential host galaxies. When the GW likelihood is instead approximated by a Gaussian distribution, thereby neglecting information from the binary black hole component masses, we obtain a consistent constraint of $H_0 = 72.6^{+11.6}_{11.3}\,\rm{km/s/Mpc}$. The two measurements are in good agreement, with the full GW likelihood yielding a slightly higher posterior maximum, primarily due to the fixed BBH mass population prior adopted in the analysis.

The inclusion of $r$-band luminosity weighting leads to percent-level changes in the $H_0$ constraints for most individual events. In the combined analysis of the 17 dark sirens, luminosity weighting broadens the 68\% CI by $\sim18\%$ when the GW likelihood is approximated by a Gaussian, but reduces it by $\sim14\%$ when the full GW likelihood is used. We also assess the impact of including uncertainties in the apparent magnitudes of galaxies and find no statistically significant effect on the results. This indicates that, at the current level of precision, these magnitudes can be safely treated as perfectly measured. The resulting luminosity-weighted constraint is $H_0 = 78.2^{+12.0}_{-11.0}~\rm{km/s/Mpc}$, consistent within uncertainties with the result obtained without luminosity weighting. This is in agreement with previous studies showing that biases from fixing a host-galaxy weighting scheme can be safely neglected for complete galaxy catalogs and well-localized events \citep{Hanselman2025, Alfradique2025, Borghi2026}. Nevertheless, this does not exclude the possibility that such choices may introduce systematic biases in other observational configurations \citep{Perna2024, Hanselman2025}.
 
Finally, we combine our 17 dark siren measurements obtained with $r$-band luminosity weighting with the bright siren GW170817, taking into account constraints from VLBI measurements of the viewing angle and corrections for the host galaxy peculiar velocity \citep{Mukherjee_2021}, to obtain a combined constraint of $H_0 = 69.9^{+4.1}_{-4.0}\,\rm{km/s/Mpc}$. This result represents an improvement of approximately 67\% and 64\% in uncertainty compared to the measurements obtained using the GW170817 bright siren result with only peculiar velocity corrections \citep{nicolaou2019impact} and using dark sirens alone, respectively. Although the bright siren provides the dominant constraint, adding the dark sirens to the GW170817 measurement still reduces the uncertainty by about 11\%, illustrating the complementarity between bright and dark sirens in GW cosmology.

The $H_0$ constraints presented here rely on fixed assumptions about the BBH merger rate and population mass distribution, adopted to enable a direct comparison with the methodology used in previous works \citep{darksiren1, palmese20_sts, Finke2021, alfradique2024, bom2024}. In particular, we model the BBH merger rate as tracing the $r$-band luminosity of galaxies, motivated by its correlation with stellar mass, and we neglect the time delay between the formation of stellar progenitors and the BBH merger. Binary population synthesis simulations combined with galaxy evolution models \citep{Artale2019a, Artale2019b, Santoliquido2021, Mandhai2022, Rauf2023} have shown, however, that the merger rate depends not only on stellar mass, but also on star formation rate and metallicity. We leave this investigation to future work. In upcoming GW observing runs, the expected increase in the number of well-localized dark siren events, together with the advent of deeper wide-field galaxy surveys such as the Vera C. Rubin Observatory Legacy Survey of Space and Time \citep{Ivezic2019}, and possible additional bright siren detections, is expected to bring GW cosmology to percent-level precision, where controlling systematic effects will become essential.

\section*{Acknowledgements}

\noindent The authors made use of Sci-Mind servers machines developed by the CBPF AI LAB team and would like to thank Paulo Russano and Marcelo Portes de Albuquerque for all the support in infrastructure matters. CRB acknowledges the financial support from CNPq (316072/2021-4) and from FAPERJ (grants 201.456/2022 and 210.330/2022) and the FINEP contract 01.22.0505.00 (ref. 1891/22). AS acknowledges the financial support from CNPq (402577/2022-1). GT aknowleges the financial support from CNPq (PhD fellowship, 140212/2022-1) and from FAPERJ (PhD merit fellowship - FAPERJ NOTA 10, 202.432/2024).

\noindent The Legacy Surveys consist of three individual and complementary projects: the Dark Energy Camera Legacy Survey (DECaLS; NSF's OIR Lab Proposal ID 2014B-0404; PIs: David Schlegel and Arjun Dey), the Beijing-Arizona Sky Survey (BASS; NSF's OIR Lab Proposal ID 2015A-0801; PIs: Zhou Xu and Xiaohui Fan), and the Mayall z-band Legacy Survey (MzLS; NSF's OIR Lab Proposal ID 2016A-0453; PI: Arjun Dey). DECaLS, BASS and MzLS together include data obtained, respectively, at the Blanco telescope, Cerro Tololo Inter-American Observatory, The NSF's National Optical-Infrared Astronomy Research Laboratory (NSF's OIR Lab); the Bok telescope, Steward Observatory, University of Arizona; and the Mayall telescope, Kitt Peak National Observatory, NSF's OIR Lab. The Legacy Surveys project is honored to be permitted to conduct astronomical research on Iolkam Du'ag (Kitt Peak), a mountain with particular significance to the Tohono O'odham Nation.

\noindent The NSF's OIR Lab is operated by the Association of Universities for Research in Astronomy (AURA) under a cooperative agreement with the National Science Foundation.

\noindent This project used data obtained with the Dark Energy Camera (DECam), which was constructed by the Dark Energy Survey (DES) collaboration. Funding for the DES Projects has been provided by the U.S. Department of Energy, the U.S. National Science Foundation, the Ministry of Science and Education of Spain, the Science and Technology Facilities Council of the United Kingdom, the Higher Education Funding Council for England, the National Center for Supercomputing Applications at the University of Illinois at Urbana-Champaign, the Kavli Institute of Cosmological Physics at the University of Chicago, Center for Cosmology and Astro-Particle Physics at the Ohio State University, the Mitchell Institute for Fundamental Physics and Astronomy at Texas A\&M University, Financiadora de Estudos e Projetos, Fundacao Carlos Chagas Filho de Amparo, Financiadora de Estudos e Projetos, Fundacao Carlos Chagas Filho de Amparo a Pesquisa do Estado do Rio de Janeiro, Conselho Nacional de Desenvolvimento Cientifico e Tecnologico and the Ministerio da Ciencia, Tecnologia e Inovacao, the Deutsche Forschungsgemeinschaft and the Collaborating Institutions in the Dark Energy Survey. The Collaborating Institutions are Argonne National Laboratory, the University of California at Santa Cruz, the University of Cambridge, Centro de Investigaciones Energeticas, Medioambientales y Tecnologicas-Madrid, the University of Chicago, University College London, the DES-Brazil Consortium, the University of Edinburgh, the Eidgenossische Technische Hochschule (ETH) Zurich, Fermi National Accelerator Laboratory, the University of Illinois at Urbana-Champaign, the Institut de Ciencies de l'Espai (IEEC/CSIC), the Institut de Fisica d'Altes Energies, Lawrence Berkeley National Laboratory, the Ludwig-Maximilians Universitat Munchen and the associated Excellence Cluster Universe, the University of Michigan, the National Optical Astronomy Observatory, the University of Nottingham, the Ohio State University, the University of Pennsylvania, the University of Portsmouth, SLAC National Accelerator Laboratory, Stanford University, the University of Sussex, and Texas A\&M University.

\noindent BASS is a key project of the Telescope Access Program (TAP), which has been funded by the National Astronomical Observatories of China, the Chinese Academy of Sciences (the Strategic Priority Research Program "The Emergence of Cosmological Structures" Grant \# XDB09000000), and the Special Fund for Astronomy from the Ministry of Finance. The BASS is also supported by the External Cooperation Program of Chinese Academy of Sciences (Grant \# 114A11KYSB20160057), and Chinese National Natural Science Foundation (Grant \# 11433005).

\noindent The Legacy Survey team makes use of data products from the Near-Earth Object Wide-field Infrared Survey Explorer (NEOWISE), which is a project of the Jet Propulsion Laboratory/California Institute of Technology. NEOWISE is funded by the National Aeronautics and Space Administration.

\noindent The Legacy Surveys imaging of the DESI footprint is supported by the Director, Office of Science, Office of High Energy Physics of the U.S. Department of Energy under Contract No. DE-AC02-05CH1123, by the National Energy Research Scientific Computing Center, a DOE Office of Science User Facility under the same contract; and by the U.S. National Science Foundation, Division of Astronomical Sciences under Contract No. AST-0950945 to NOAO.

\noindent The Photometric Redshifts for the Legacy Surveys (PRLS) catalog used in this paper was produced thanks to funding from the U.S. Department of Energy Office of Science, Office of High Energy Physics via grant DE-SC0007914.

\section*{Data Availability}
 
The data underlying this article will be shared on reasonable request to the corresponding author.

\appendix

\section{Effects of photo-$z$ non-Gaussianity and luminosity weighting on individual $H_0$ posteriors}\label{app:lum_eff}

In this appendix, we examine the impact of approximating galaxy redshift measurements with a Gaussian distribution on the inferred $H_0$ posteriors. We also analyze how applying \textit{r}-band luminosity weighting modifies the redshift distributions of potential host galaxies and propagates into the $H_0$ inference for individual dark siren events.

In most dark siren analyses, galaxy photometric redshift uncertainties are modeled using a Gaussian approximation \citep[see, e.g.,][]{darksiren1, Finke2021, Abbott_2023, gwtc4_cosmo}. As shown by \citet{Turski2023}, neglecting redshift uncertainties can introduce a bias comparable to other known systematic effects, although still much smaller than the dominant statistical uncertainty in GWTC-3 constraints. Consequently, the choice of redshift error model does not significantly affect the current precision of the Hubble constant measurement. However, the use of the full photo-$z$ probability density functions (here estimated with the MDN technique; see section~\ref{sec:photoz}) offers an important advantage: by naturally capturing the multiple peak structures arising from color–redshift degeneracies, it yields a more accurate description of small scale features in the redshift distribution $N(z)$, as shown by \citet{palmese20_sts}, and as already established in several studies of photometric survey analyses \citep{Oyaizu_2008, Bonnett_2016, Tanaka_2017}. Another important aspect is that the Gaussian photo-$z$ approximation also introduces a dependence on the redshift cuts applied to the volume-limited sample, due to the non-negligible contribution of its tails at high redshift (and hence high $H_0$), as discussed by \citet{darksiren1} and evidenced in \citet{palmese20_sts}. Although systematic effects associated with the Gaussian photo-$z$ approximation have not yet been widely explored, their impact may become more important as the precision of dark siren measurements improves. As galaxy catalogs become more complete and the number of GW detections increases, reducing the statistical errors, such systematics could become statistically relevant for $H_0$ inference.

\begin{figure*}
\centering
\includegraphics[width=0.94\linewidth]{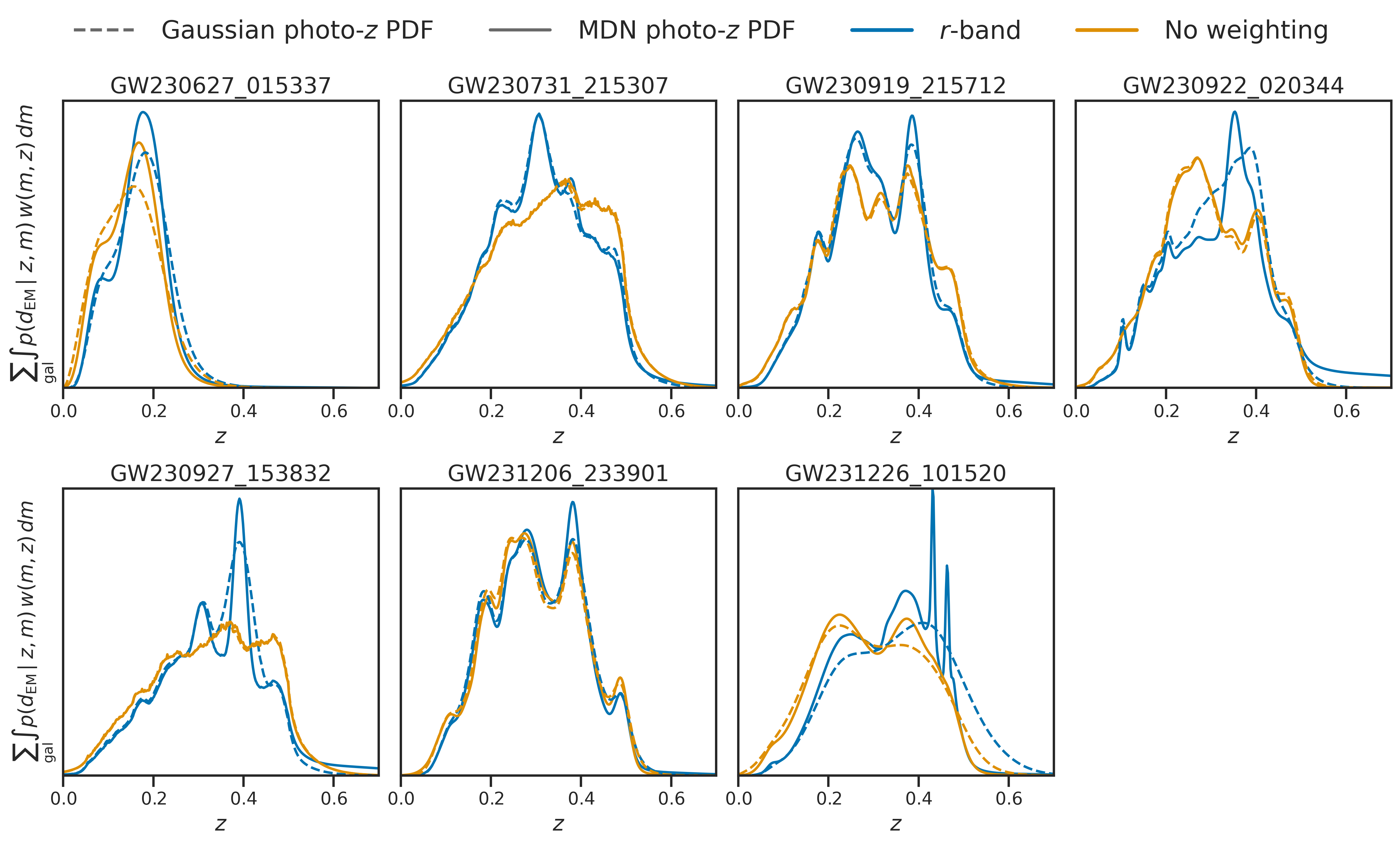}
\caption{Redshift distributions of possible host galaxies for individual GW events. Each panel shows the sum over galaxies of the integrated product of the electromagnetic likelihood and the galaxy host weighting as a function of redshift $z$. Solid lines correspond to the photo-$z$ PDFs obtained with the MDN method, while dashed lines correspond to the Gaussian approximation. The colors indicate the weighting scheme: $r$-band luminosity weighting (blue) and unweighted (orange). The distributions are normalized to unity in each panel for better visual comparison.}
\label{fig:pem_weight_int}
\end{figure*}

We focus on comparing the $H_0$ posterior distributions obtained using the full photo-$z$ probability density functions estimated with the MDN approach adopted in this work with those based on a Gaussian approximation. In the latter case, each galaxy photo-$z$ PDF is replaced by a Gaussian distribution whose mean and standard deviation are set to match the peak and the width of the corresponding full photo-$z$ PDF, respectively. We explore this comparison across all configurations adopted in this work: (i) equal versus $r$-band luminosity weighting, and (ii) the full versus Gaussian implementations of the GW likelihood. Figure~\ref{fig:pem_weight_int} shows the redshift distributions of potential host galaxies obtained using either the full MDN photo-$z$ PDFs (solid lines) or their Gaussian approximations (dashed lines). For most configurations in the no-weighting case, the Gaussian approximation produces no statistically significant change in the shape of the resulting $H_0$ posteriors, with the solid and dashed orange curves nearly overlapping. An exception occurs for GW230627\_015337 and GW231226\_101520, where the Gaussian approximation yields a slightly more precise $H_0$ constraint and introduces a small shift in its peak, both below the 1\% level.

The greatest impact occurs when $r$-band luminosity weighting is applied, with the strongest effects observed for GW230627\_015337, GW230922\_020344, and GW231226\_101220, as evident in Fig.~\ref{fig:pem_weight_int}. The Gaussian approximation leads to noticeable shifts in the credible intervals, which propagate directly to the inferred $H_0$ posterior. For GW230922\_020344, it shifts the posterior peak from $20.0~\mathrm{km/s/Mpc}$ to $68.5~\mathrm{km/s/Mpc}$, with a $\sim$5\% reduction in the 68\% CI when the GW likelihood is approximated as Gaussian. An even larger effect is observed for GW230627\_015337, where the posterior mode shifts by $\sim$91(47) km/s/Mpc for the Gaussian (full) GW likelihood case, while the uncertainty decreases by 5\%(6\%). In GW231226\_101220, the approximation instead modifies the $H_0$ uncertainty, increasing (decreasing) it by about 5\% in the Gaussian (full) GW likelihood case. These deviations highlight the importance of accounting for non-Gaussian features in the photo-$z$ PDFs and indicate that Gaussian approximations can introduce systematic biases in individual dark siren $H_0$ inferences. This motivates further investigation with realistic simulations and different observing scenarios for galaxy surveys and GW detections, which we plan to explore in future work.

\begin{figure}
    \centering
    \includegraphics[width=\linewidth]{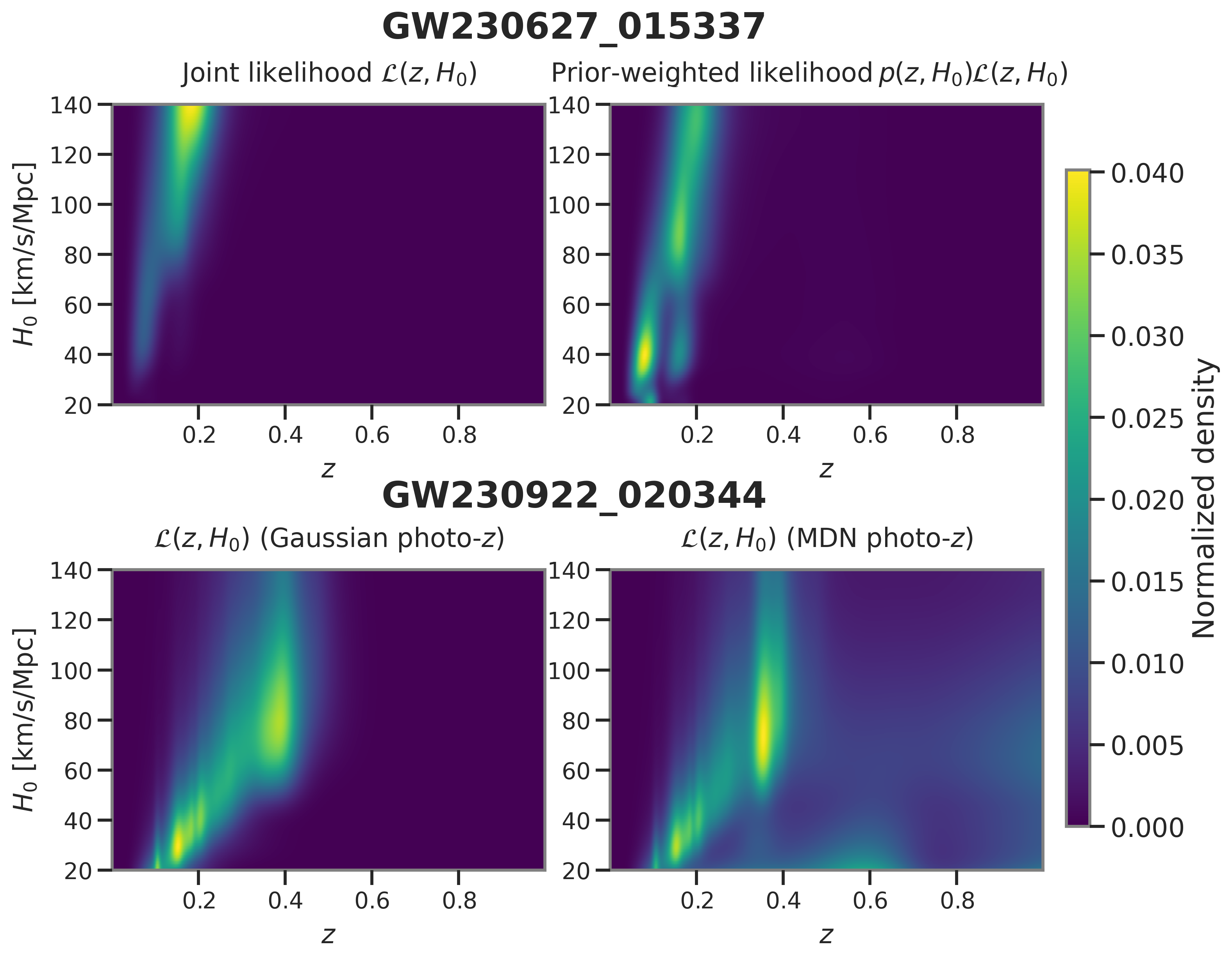}
    \caption{Two–dimensional joint GW–EM likelihood in the $(z,H_0)$ plane for GW230627\_015337 (top row) and GW230922\_020344 (bottom row). In the top row, the left panel shows the joint likelihood $\mathcal{L}(z,H_0)$, while the right panel shows the integrand of the marginalized likelihood obtained when adopting a redshift prior uniform in comoving volume ($p(z,H_0)\propto dV/dz$). The bottom row displays the likelihood for GW230922\_020344 computed using two different photo-$z$ PDF treatments: a Gaussian approximation (left) and the full photo-$z$ PDF predicted by the MDN (right). Colors represent the normalized probability density.}
    \label{fig:lum_eff_app}
\end{figure}

We now examine the effect of $r$-band luminosity weighting on the host galaxy redshift distributions and the resulting $H_0$ inference. For GW230627\_015337, we find that luminosity weighting broadens the redshift distribution of potential host galaxies, which in turn flattens the joint GW–EM likelihood and increases the relative contribution of the redshift prior in the marginalization over $z$. When marginalized over redshift, this term peaks at low $H_0$ because the comoving volume element scales as $dV/dz \propto H_0^{-3}$ at fixed $z$. At low $H_0$, the likelihood still retains non-negligible support, but the redshift prior increases the relative weight of these regions in the marginalization, driving the inference toward smaller $H_0$ values. The effect becomes more pronounced when luminosity weighting is applied, as it broadens the joint GW–EM likelihood for GW230627\_015337, reducing the information gain relative to the prior from $D_{\rm KL}=1.02$ to $0.75$. This behavior is illustrated in the top panel of Fig.~\ref{fig:lum_eff_app}. The left panel shows the joint GW–EM likelihood in the $(z,H_0)$ plane, which is not maximal at low $H_0$ (e.g., $H_0 \simeq 40~\mathrm{km/s/Mpc}$) but still retains non-negligible support. The right panel shows the same likelihood weighted by the redshift prior $p(z, H_0)\propto dV/dz$, clearly enhancing the probability toward smaller values of $H_0$.

For GW230922\_015337, using the Gaussian approximation to the GW likelihood (dashed lines in Fig.~\ref{fig:pH0_all}), we find that the $H_0$ posterior in the no-weighting case peaks at $H_0 \simeq 50~\mathrm{km/s/Mpc}$. In contrast, the luminosity-weighted case shows a broader distribution with a secondary peak near $H_0 \simeq 68\,\mathrm{km/s/Mpc}$, but with higher probability density toward lower values around $H_0 \simeq 20\,\mathrm{km/s/Mpc}$. This shift is driven by the change in the redshift distribution of potential host galaxies: the no–weighting case (orange curve in Fig.~\ref{fig:pem_weight_int}) peaks at $z \simeq 0.3$, whereas the luminosity–weighting case (blue curve) peaks at $z \simeq 0.4$. The extended tail toward low $H_0$ arises from the high–$z$ tail of the photometric redshift PDFs modeled with the MDN methodology. The non-Gaussian features in the photo–$z$ PDFs significantly affect the shape of the redshift distribution along the line of sight in the luminosity–weighting scenario. This is illustrated in the bottom panel of Fig.~\ref{fig:lum_eff_app}, where the MDN–based case (right) shows non-zero probability at $H_0 \simeq 20\,\mathrm{km/s/Mpc}$ for high $z$, in contrast to the Gaussian photo–$z$ approximation (left).

\bibliography{references_prd}

\end{document}